\documentclass[prb,aps,nobibnotes,twocolumn]{revtex4}
\usepackage{graphicx}%
\usepackage{dcolumn}
\usepackage{amsmath}   
\usepackage{bm}
\begin{document}

\title{\centering\Large\bf Equilibrium solvation in quadrupolar solvents}
\author{Anatoli A.\ Milischuk and Dmitry V.\ Matyushov}
\email[E-mail:]{dmitrym@asu.edu.}
\affiliation{Department of Chemistry
and Biochemistry, Arizona State University , PO Box 871604, Tempe, 
AZ 85287-1604}
\date{\today}
\begin{abstract}
We present a microscopic theory of equilibrium solvation in solvents
with zero dipole moment and non-zero quadrupole moment (quadrupolar
solvents).  The theory is formulated in terms of autocorrelation
functions of the quadrupolar polarization (structure factors). It can
be therefore applied to an arbitrary dense quadrupolar solvent for which the
structure factors are defined. We formulate a simple analytical
perturbation treatment for the structure factors. The solute is
described by coordinates, radii, and partial charges of constituent
atoms. The theory is tested on Monte Carlo simulations of solvation in
model quadrupolar solvents. It is also applied to the calculation of
the activation barrier of electron transfer reactions in a
cleft-shaped donor-acceptor complex dissolved in benzene with the structure
factors of quadrupolar polarization obtained from Molecular Dynamics simulations.
\end{abstract}

\preprint{Submitted to J.\ Chem.\ Phys.}
\maketitle

\section{Introduction}
\label{sec:0}
The understanding of kinetics of chemical reactions in non-polar or,
more generally, non-dipolar solvents poses the necessity to describe
thermodynamic\cite{DMjcp:95,Reynolds:96,Ladanyi:98,Larsen:99,DMjcp2:99,Raineri:99}
and dynamic\cite{Stephens:97,Berg:98,BagchiBiswas:99,YamaguchiJPC:02,Egorov:02} aspects of
non-polar solvation.\cite{comQ:1} The overall solvation free energy in
a non-dipolar solvent (zero permanent dipole) can be approximately
separated into the contribution from the solute repulsive core
(cavitation energy), the attractive dispersion solvation, and
electrostatic contributions from induced dipoles and permanent
multipoles starting from the quadrupole moment. The cavitation energy
is often described by models considering the free energy necessary to
insert the solute hard repulsive core into the
solvent.\cite{Reiss:59,Stamatopoulou:97,DMjcp3:97,Ben-Amotz:01,Heying:04}
The dispersion energy is a very significant part of solvation
energetics even in polar solvents.\cite{DMjpcb:97} It is normally
modeled by site-site Lennard-Jones (LJ) interaction potentials,
although LJ parameterization is often obscure for molecular excited states.

The positive cavitation free energy and negative contributions from
dispersion and electrostatic interactions cancel each other in the
overall solvation free energy which often constitutes only a small
portion of each component. The cavitation and dispersion energies,
however, cancel almost identically when the absorption and emission
solvatochromic shifts are subtracted to form the optical Stokes shift
or when the solvent reorganization energy of electron transfer (ET) is
calculated.\cite{DMjcp:95,DMjpcb:97} The remaining contribution from
dispersion interactions, normally associated with mechanical
solvation,\cite{Fourkas:94,Tran:99} does not typically exceed
$100-200$ cm$^{-1}$, which is small compared to usual values of the
Stokes shift arising from dipolar and quadrupolar
solvation.\cite{Reynolds:96} The electrostatic component of solvation
free energy in non-dipolar solvents arising from (partial) solute
charges interacting with solvent quadrupoles is the focus of the
present equilibrium solvation theory.

The electrostatic component of non-dipolar solvation, in particular
quadrupolar solvation, has received little
attention\cite{Reynolds:96,Larsen:99,Ladanyi:98,Raineri:99,Jeon:01,Kim1:03,Kim2:03,Nugent:04}
compared to the very extensive literature on solvation in dipolar
solvents.\cite{Barbara:90,Bagchi:91,Raineri:99} In contrast to dipolar
solvation, where dielectric measurements provide the basis for
modeling thermodynamics and dynamics of
solvation,\cite{Zwan:85,Fried:90,Barbara:90,Maroncelli:93,Horng:95,Hsu:97}
there is no obvious method to extract the dynamic and thermodynamic
response functions in non-dipolar solvents from existing data. This
paper aims at bridging this gap by formulating a microscopic solvation
theory in terms of correlation functions of quadrupolar polarization
of the pure solvent.  In this paper, we limit the solvent correlation
functions to the spatial domain thus gaining access to the equilibrium
solvation thermodynamics. Once dynamic correlation functions become
available, the theory can be extended to solvation dynamics.

The experimental evidence on electrostatic quadrupolar solvation comes
from the realm of steady-state\cite{Morais:95,Reynolds:96,Kauffman:01} and
time-resolved\cite{Reynolds:96,Larsen:99} optical spectroscopy,
resonance Raman spectroscopy,\cite{Britt:95,Kulinowski:95} and from the
kinetics of electron transfer (ET) reactions.\cite{ZimmtWaldeck:03}
Spectroscopic chromophores and ET solutes are normally large molecules
with complex molecular shape.  Because of the relatively short range
of interaction of molecular solvent quadrupoles with solute partial
charges, it might be critical to include the correct molecular shape of
the solute into the formalism. We therefore sacrifice some
accuracy of the modeling compared to potentially more accurate
perturbation models for spherical solutes\cite{DMjcp2:99} in order to
introduce the molecular shape of the solute with atomic resolution
combined with molecular charge distribution specified by atomic
partial charges.  It appears that this is the first theory of
quadrupolar solvation approaching this level of detail in describing
the solute.

 \begin{figure}[tbh]
 \includegraphics[width=8cm]{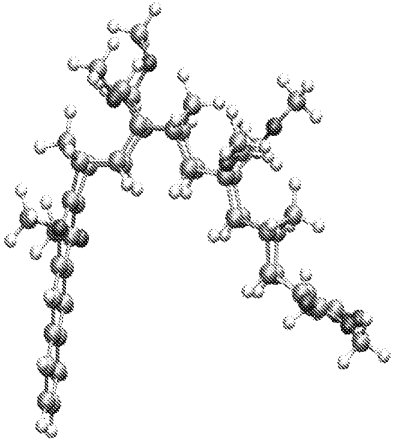}
   \caption{Charge transfer complex combining dymethoxyanthracene unit for the donor (D) and 
     a cyclobutene dicarboxylate derivative for an acceptor (A)
     connected by a bridge (B).\cite{ZimmtWaldeck:03} The present
     theory is applied to the activated kinetics of charge separation,
     DB$^*$A$\to$D$^+$BA$^-$.}
   \label{fig:1}
\end{figure}

The paper starts with the formulation of the problem (Sec.\ 
\ref{sec:1}) followed by the formal theory of solvation thermodynamics
(Sec.\ \ref{sec:2}). The perturbation theory is given in terms of
structure factors of fluctuating quadrupolar polarization in the pure
solvent discussed in Sec.\ \ref{sec:3}. The theory is tested on
computer simulations of model ionic and dipolar solutes in Sec.\
\ref{sec:4} and is compared to experimental ET kinetics in Sec.\ 
\ref{sec:5}. ET kinetics in a donor-bridge-acceptor cleft molecule
referred to as complex \textbf{1} (Fig.\ \ref{fig:1}) has been
extensively studied by Waldeck and Zimmt.\cite{ZimmtWaldeck:03} The
present theory is applied to the kinetic data in benzene used as a
solvent. As a consistency test, the parameters used for ET rates in
benzene are applied to ET in acetonitrile with the previously
developed theory of polar solvation.\cite{DMjcp2:04,DMcp:05} The
overall good agreement between theory and experiment is reported.

\section{Conceptual framework}
\label{sec:1}
Solvation in polar solvents is defined by the coupling of the field
$\mathbf{E}_0(\mathbf{r})$ of the solute to the dipolar polarization
of the solvent $\mathbf{P}(\mathbf{r})$. The interaction potential of
the solute (subscript ``0'') with the solvent (subscript ``s'') is then
a composite effect of this coupling integrated over the space occupied
by the solvent $\Omega$
\begin{equation}
  \label{eq:1-1}
  v_{0s}[\mathbf{P}] = - \int_{\Omega} \mathbf{P}(\mathbf{r})\cdot \mathbf{E}_0(\mathbf{r}) d\mathbf{r}.
\end{equation}
Here, the dipolar polarization is defined by the density of permanent dipoles
$\mathbf{m}_j$ in the liquid
\begin{equation}
  \label{eq:1-2}
  \mathbf{P}(\mathbf{r}) = \sum_j \mathbf{m}_j \delta(\mathbf{r}-\mathbf{r}_j) ,
\end{equation}
where the sum runs over $N$ molecules of the solvent with
center-of-mass coordinates $\mathbf{r}_j$.  In the linear response
approximation (LRA), the above interaction energy is supplemented by
the Gaussian Hamiltonian\cite{Chandler:93}
\begin{equation}
  \label{eq:1-3}
  H_P[\mathbf{P}] = \frac{1}{2} \int \mathbf{P}(\mathbf{r})\cdot 
       \bm{\chi}_P(\mathbf{r},\mathbf{r}')^{-1}\cdot \mathbf{P}(\mathbf{r}') d\mathbf{r} d\mathbf{r}',
\end{equation}
where the polarization response function
$\bm{\chi}_P(\mathbf{r},\mathbf{r}')$ generally depends on the shape of
the field source. For certain geometries, e.g.\ for a parallel plate
capacitor, the dependence on geometry can be eliminated. The
polarization induced in the solvent by the external electric field
$\mathbf{E}_0(\mathbf{r})$ is then obtained by minimizing the
functional $v_{0s}[\mathbf{P}] + H_P[\mathbf{P}]$ in
$\mathbf{P}(\mathbf{r})$ to yield
\begin{equation}
  \label{eq:1-4}
  \mathbf{P}(\mathbf{r}) = \int_{\Omega} \bm{\chi}_P(\mathbf{r},\mathbf{r}')\cdot\mathbf{E}_0(\mathbf{r}')d\mathbf{r}'.
\end{equation}

Molecular quadrupoles couple to an inhomogeneous electric field with
the gradient $\nabla \mathbf{E}_0(\mathbf{r})$
\begin{equation}
  \label{eq:1-5}
  v_{0s}[\mathbf{Q}] = - (1/3) \int_{\Omega} \mathbf{Q}(\mathbf{r}): \nabla \mathbf{E}_0(\mathbf{r}) d\mathbf{r},
\end{equation}
where the quadrupolar polarization is 
\begin{equation}
  \label{eq:1-6}
  \mathbf{Q}(\mathbf{r}) = \sum_j \mathbf{Q}_j \delta(\mathbf{r} - \mathbf{r}_j) 
\end{equation}
and $\mathbf{Q}_j$ is the molecular quadrupole tensor. Similarly to Eq.\ 
(\ref{eq:1-3}), the Hamiltonian of the quadrupolar polarization in the
pure solvent is given in the Gaussian form
\begin{equation}
  \label{eq:1-7}
  H_Q[\mathbf{Q}] = \frac{1}{2} \int_{\Omega} \mathbf{Q}(\mathbf{r}):
                   \bm{\chi}_Q(\mathbf{r},\mathbf{r}')^{-1}:\mathbf{Q}(\mathbf{r}')d\mathbf{r}d\mathbf{r}'.
\end{equation}
The minimization of $v_{0s}[\mathbf{Q}]+H_Q[\mathbf{Q}]$ in terms of
$\mathbf{Q}(\mathbf{r})$ then leads to 
\begin{equation}
  \label{eq:1-8}
  \mathbf{Q}(\mathbf{r}) = \frac{1}{3}\int_{\Omega} \bm{\chi}_Q(\mathbf{r},\mathbf{r}') : \nabla \mathbf{E}_0(\mathbf{r}') d\mathbf{r}' .
\end{equation}

The definition of response functions $\bm{\chi}_P(\mathbf{r},\mathbf{r}')$
and $\bm{\chi}_Q(\mathbf{r},\mathbf{r}')$ incorporates the non-local
solvent response affected by finite-range microscopic correlations of
molecular dipoles and quadrupoles. These correlations are neglected in
the continuum approximation
\begin{equation}
  \label{eq:1-9}
  \bm{\chi}_{P,Q}(\mathbf{r},\mathbf{r}') = \delta(\mathbf{r} - \mathbf{r}')\bm{\chi}_{P,Q}(\mathbf{r}).
\end{equation}
The continuum approximation, used in dielectric continuum models of
dipolar\cite{Onsager:36} and non-dipolar\cite{Kim1:03,Kim2:03} solvation,
significantly simplifies the calculation of the polar response
function. In particular, the dipolar response function can be related
to the macroscopic dielectric properties of the solvent through the
macroscopic material Maxwell's equations.

The Maxwell's equation for the overall electric field in
the dielectric $\mathbf{E}$ reads\cite{Boettcher:73}
\begin{equation}
  \label{eq:1-10}
  \nabla\cdot\,\left(\mathbf{E} + 4\pi \mathbf{P} \right) = 4\pi\rho + 4\pi 
                \nabla\cdot\,(\nabla\cdot\,\mathbf{Q}) .
\end{equation}
The material equations are closed by defining the dielectric
displacement $\mathbf{D} = \mathbf{E} + 4\pi\mathbf{P}$ which is
connected to the density of external charges $\rho$ by the relation
neglecting the quadrupolar density in the right hand part of Eq.\ 
(\ref{eq:1-10})
\begin{equation}
\nabla\cdot\, \mathbf{D} = 4\pi\rho . 
\end{equation}
The dielectric displacement is in turn related to the
overall electric field through the static dielectric constant $\epsilon_s$,
$\mathbf{D}=\epsilon_s \mathbf{E}$. In the case of a parallel plate capacitor,
$\mathbf{D}=\mathbf{E}_0$ and one gets
\begin{equation}
  \label{eq:1-11}
  \chi_P = (\epsilon_s - 1) / 4\pi\epsilon_s .
\end{equation}
More complex geometries of the field source require solving the
Poisson equation with the boundary conditions defined by the
dielectric constant and the shape of the dielectric. This procedure
establishes the widely used dielectric continuum approximation for the
dipolar response.

Many of the advantages of the dielectric continuum approximation
disappear when applied to quadrupolar (non-dipolar) solvation. The
main problem is that the continuum quadrupolar susceptibility
$\bm{\chi}_Q$ does not come to the material Maxwell's equations and is,
therefore, not directly related to any well-established experimental
protocol. Measuring quadrupolar susceptibility is still a non-trivial
experimental problem.\cite{Ernst:92} In practical continuum
calculations of the quadrupolar response, the quadrupolar
susceptibility is obtained by fitting the calculated response to
solvation free energies from spectroscopy.\cite{Kim2:03} One would
alternatively want to have the quadrupolar susceptibility from
properties of a pure quadrupolar solvent unaffected by all the
complexities of treating spectral band-shapes of complex molecular
solutes. In addition, one needs to know the limits of applicability of
the continuum approximation [Eq.\ (\ref{eq:1-9})] to relatively
short-ranged quadrupolar interactions. The high directionality of
quadrupolar forces make them unlike candidates for mean-field theories
which are successfully applied to short-range but more isotropic
dispersion forces.\cite{Widom:67,Gelbart:77} All these considerations
call for the necessity of microscopic theories of quadrupolar
solvation. Once such a theory is formulated, approximate solutions,
e.g.\ the continuum limit, can be obtained in a more controlled
fashion. The formulation of such a theory is a goal of this paper.

In order to develop a theory applicable to solutes of complex shape,
we use a particular expression of the LRA to obtain the chemical
potential of solvation $\mu_{0s}$. Within the LRA, $\mu_{0s}$ can be
calculated as the second cumulant of the solute-solvent interaction
potential\cite{DMjpca:02}
\begin{equation}
  \label{eq:1-12}
  -\mu_{0s}= (\beta/2)\left\langle \delta v_{0s}^2 \right\rangle_0,
\end{equation}
where $\beta=1/k_{\text{B}}T$, $k_{\text{B}}$ is Boltzmann's constant and
$T$ is the temperature. The statistical average $\left\langle\dots
\right\rangle_0$ is over the solvent configurations around a fictitious
solute with the solute-solvent interaction energy $v_{0s}$ eliminated
from the total interaction energy. The calculation of the average
$\langle\dots\rangle_0$ over the solvent configurations in the presence of the
repulsive core of the solute is a major challenge for the theory
development.  The solute expels the solvent from its volume and may
significantly modify the statistics of solvent fluctuations either by
altering the local density profile of the solvent or/and the
statistics of orientational fluctuations of the solvent
molecules.\cite{Chandler:93} The expulsion effect propagates through
the entire solvent changing substantially the solvent response
function in strongly dipolar liquids.\cite{Song:98,DMjcp1:04} This
strong effect of the solute on solvent statistics arises from the long
range character of dipole-dipole interactions accounted for in
dielectric models through the boundary conditions in the Poisson
equation.

The quadrupole-quadrupole interaction is much more short-range
compared to the dipole-dipole interaction ($\propto1/r^5$ vs $\propto1/r^3$). One
may expect that statistics of orientational quadrupolar fluctuations
is not significantly altered by the solute which only expels the
solvent quadrupoles from its volume.  In this approximation, one can
switch from the statistical average $\langle\dots\rangle_0$ to the average
$\langle\dots\rangle$ over the statistical configurations of the pure solvent by
replacing $v_{0s}(\mathbf{r})$ in Eq.\ (\ref{eq:1-12}) with
$v_{0s}(\mathbf{r})\theta(\mathbf{r})$, where $\theta(\mathbf{r})$ is a step
function equal to zero inside the solute and equal to one everywhere
else.

We will denote the component of the solvation chemical potential
arising from fluctuations of quadrupolar polarization in homogeneous
solvent as $\mu_{0s}^Q$ (superscript ``Q'' stands for quadrupolar
polarization). This component is expected to be the major contribution
to the quadrupolar solvation chemical potential. However, in addition
to expelling quadrupolar polarization from its volume, insertion of a
solute into a liquid creates a nonuniform density profile represented
by the solute-solvent pair correlation function $h_{0s}(\mathbf{r})$.
This correlation function, highly specific to the solute shape and the
thermodynamic state of the solvent, can be reliably calculated only
for simple solute geometries. The component of $\mu_{0s}$ associated
with $h_{0s}(\mathbf{r})$ will be denoted as $\mu_{0s}^D$ (superscript
``D'' refers to the local density profile).  The overall chemical
potential of solvation is a sum of the long-range component due to
quadrupolar orientational fluctuations (``Q'' component) and the
short-range component due to the local density profile (``D''
component):
\begin{equation}
  \label{eq:1-13}
  \mu_{0s} = \mu_{0s}^Q + \mu_{0s}^D .
\end{equation}

\section{Perturbation theory of solvation}
\label{sec:2}
The solute-solvent interaction potential in Eq.\ (\ref{eq:1-5}) can be
written either in Cartesian or spherical coordinates. We will use
Greek indexes for the Cartesian projections and Latin indexes for the
spherical projections.\cite{Gubbins:84} In the Cartesian projections,
the solute electric field gradient and the quadrupolar polarization are given as
follows
\begin{equation}
\label{eq:2-2}
\phi_{\alpha\beta}(\mathbf{r})=-\nabla_{\alpha}E_{\beta}=-\nabla_{\alpha}\nabla_{\beta}\sum_{a=1}^M\dfrac{q_0^a}{\left|\mathbf{r}-\mathbf{r}_0^a\right|} 
\end{equation}
and
\begin{equation}
\label{eq:2-3}
Q_{\alpha\beta}( \mathbf{r}) =\sum_j Q_{\alpha\beta,j}\delta(\mathbf{r}-\mathbf{r}_j) .
\end{equation}  
The sum (index $a$) in Eq.\ (\ref{eq:2-2}) runs over the $M$ solute 
(subscript ``0'') atoms
with coordinates $\mathbf{r}_0^a$ and partial charges $q_0^a$. In Eq.\ 
(\ref{eq:2-3}), the summation is over $N$ solvent molecules with
centers of mass at $\mathbf{r}_j$ relative to which the quadrupole
tensor $Q_{\alpha\beta,j}$ is defined as:
\begin{equation}
\label{eq:2-4}
    Q_{\alpha\beta,j}=(1/2)\sum_{a=1}^K q_{j}^a(r^a_j)^2(3 \hat r_{\alpha,j}^{a} \hat r_{\beta,j}^{a} -\delta_{\alpha\beta}) .
\end{equation}
Here, the sum (index $a$) runs over the $K$ atoms of the solvent
molecule with coordinates $\mathbf{r}_j+\mathbf{\hat r}_j^a r_j^a$
($\mathbf{\hat r}_j^a = \mathbf{r}_j^a/r_j^a$) and partial charges
$q_j^a$. 

\begin{figure}[tbh]
  \includegraphics[width=6cm]{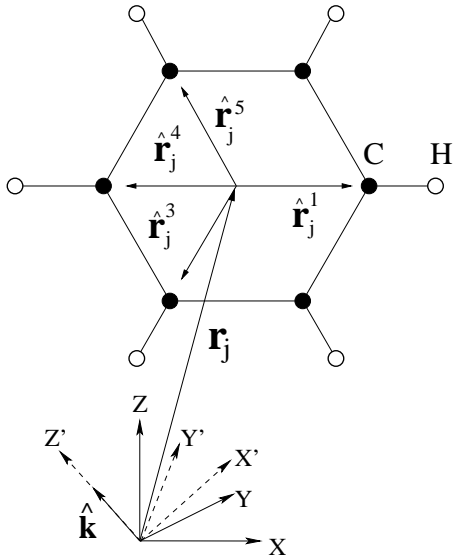}
   \caption{The schematic representation of the benzene molecule: $\mathbf{r}_j$ is 
     the coordinate of the center of mass, $\mathbf{\hat r}_{j}^a$ are
     unit vectors in the direction of partial charges $q_j^a$. $XYZ$ is the laboratory coordinate frame,
     the $Z'$ axis of the laboratory system $X'Y'Z'$ is along the unit wavevector 
     $\mathbf{\hat k} = \mathbf{k}/k$. }
   \label{fig:2}
\end{figure}

In the spherical projections one gets
\begin{equation}
\label{eq:2-6}
v_{0s}=\sum_{m,j}\int  \phi_{2m}(\mathbf{r})Q^*_{2m,j}(\mathbf{r})d\mathbf{r},
\end{equation} 
where the spherical projections of the quadrupolar polarization 
\begin{equation}
\label{eq:2-7}
Q_{2m}(\mathbf{r})=\sum_{j}Q_{2m,j}\delta(\mathbf{r}-\mathbf{r}_j)
\end{equation}
are given in terms of the spherical quadrupolar tensor
\begin{equation}
\label{eq:2-8}
     Q_{2m,j}=\sum_{a=1}^Kq_{j}^a (r_j^a)^2 Y_{2m}(\mathbf{\hat r}^{a}_j) .
\end{equation}
In Eq.\ (\ref{eq:2-8}), $Y_{2m}(\mathbf{\hat r})$ is the spherical
harmonic of the second order, $-2\leq m\leq2$. The distribution of molecular charge is
illustrated in Fig.\ \ref{fig:2} on the example of the benzene
molecule the first non-zero multipole of which is quadrupole.

Substituting $v_{0s}$ from Eq.\ (\ref{eq:2-6}) into Eq.\ (\ref{eq:1-12})
and switching to the $\mathbf{k}$-space we obtain
\begin{equation}
\label{eq:2-10}
\mu_{0s}^Q =-\dfrac{\beta \rho Q^2}{8\pi}\sum_{m,n}\int\dfrac{d\mathbf{k}}{\left(2\pi \right)^3}
          \tilde\phi_{2m}\left(-\mathbf{k} \right) \tilde\phi^*_{2n}\left(\mathbf{k} \right)S_{mn}(k) .
\end{equation}
Here, $\tilde\phi_{2m}$ is the Fourier transform of $\phi_{2m}$, $\tilde\phi^*_{2n}$ is the Fourier transform of $\phi^*_{2n}$ , and
\begin{equation}
\label{eq:Q}
       Q^2 = (2/3)\mathbf{Q}:\mathbf{Q} 
\end{equation}
is the rotational-invariant ``effective axial quadrupole
moment''.\cite{Gubbins:84} The correlation function $S_{mn}(k)$ in
Eq.\ (\ref{eq:2-10}) does not depend on the orientation of the
wavevector because of the rotational isotropy of the solvent:
\begin{equation}
\label{eq:2-11}
S_{mn}(k)=\dfrac{4\pi}{NQ^2}\left\langle \sum_{i,j} Q_{2m,i}^* Q_{2n,j}  
e^{ i \mathbf{k} \cdot \mathbf{r}_{ij}}  \right\rangle.
\end{equation}

Since $\mu_{0s}^Q$ is invariant with respect to rotations of the
coordinate system we first consider $\tilde \phi_{2m}$ and $S_{mn}$ in
the coordinate system $X'Y'Z'$ in which the wavevector $\mathbf{k}$ is
directed along the $Z'$-axis (Fig.\ \ref{fig:2}). The functions of
$\mathbf{k}$ in this coordinate system will be specified with the
prime.  The wavevector $\mathbf{k}$ introduces axial symmetry in the
otherwise isotropic liquid of solvent molecules. The operation of
statistical average must therefore commute with the operation of
rotation about the wavevector $\mathbf{k}$
\begin{equation}
\label{eq:2-12}
\exp\left(-i\hat l_z\gamma \right)S'_{mn}\exp\left(i\hat l_z\gamma \right)=\exp\left( i(m-n)\gamma \right) S'_{mn},
\end{equation}
where $\hat l_z$ is the operator of rotation through the angle $\gamma$.  The
condition of invariance requires $m=n$ leading to a simplified form of
Eq.\ (\ref{eq:2-10})
\begin{equation}
\label{eq:2-13}
\mu_{0s}^Q = -\dfrac{\beta \rho Q^2}{8\pi}\sum_{m=0}^2 \int\dfrac{d\mathbf{k}}{\left(2\pi \right)^3} \tilde{\phi'}^m\left(\mathbf{k} \right){S'}^{m}(k).
\end{equation}
In Eq.\ (\ref{eq:2-13}) we represent ${\tilde \phi}'^m\left(\mathbf{k}\right)$ as
\begin{equation}
\label{eq:2-13a}
\begin{split}
{\tilde \phi}'^0\left(\mathbf{k}\right)&={\tilde \phi}'_0\left(\mathbf{k}\right),\\
{\tilde \phi}'^1\left(\mathbf{k}\right)&=\dfrac{1}{2}\bigl({\tilde \phi}'_1\left(\mathbf{k}\right)+{\tilde \phi}'_{\underline{1}}\left(\mathbf{k}\right)\bigr),\\
{\tilde \phi}'^2\left(\mathbf{k}\right)&=\dfrac{1}{2}\bigl({\tilde \phi}'_2\left(\mathbf{k}\right)+{\tilde \phi}'_{\underline{2}}\left(\mathbf{k}\right)\bigr),
\end{split}
\end{equation}
where  ($\underline{m}=-m$) and 
\begin{equation}
\label{eq:2-13b}
{\tilde \phi}'_m\left(\mathbf{k}\right)=\dfrac{\tilde\phi_{2m}(-\mathbf{k}) \tilde{\phi^*}_{2m}(\mathbf{k}) }{4\pi} .
\end{equation}

Quadrupolar structure factors  ${S'}^m(k)$ in Eq.\ (\ref{eq:2-13}) are defined as
\begin{equation}
\label{eq:2-14a}
\begin{split}
{S'}^0(k)&=S'_0(k),\\
S'^1(k)&=S'_1(k)+S'_{\underline{1}}(k),\\
S'^2(k)&=S'_2(k)+S'_{\underline{2}}(k),
\end{split}
\end{equation}
where
\begin{equation}
\label{eq:2-14}
S'_{m}(k)=\dfrac{4\pi}{NQ^2}\left\langle \sum_{i,j} (Q')^*_{2m,i} (Q')_{2m,j}  e^{i \mathbf{k}\cdot \mathbf{r}_{ij}} \right\rangle.
\end{equation}

Three quadrupolar structure factors $S'^m(k)$, grouped according to 
rotational symmetry, form a minimal set of correlation functions
describing collective fluctuations of the quadrupolar polarization.
Note that two structure factors representing uncoupled longitudinal
and transverse dipolar polarization are sufficient to describe
orientational fluctuations in dipolar
solvents.\cite{Madden:84,DMjcp2:04} Numerical calculations are more
convenient to carry out in Cartesian coordinates in which the chemical
potential of solvation and the structure factors can be re-written as
(Appendix \ref{0})
\begin{equation}
\label{eq:2-20}
           \mu_{0s}^Q = -\dfrac{\beta \rho Q^2}{2}\sum_{m=0}^2\int\dfrac{d\mathbf{k}}{\left(2\pi \right)^3} 
                   \tilde \phi^{m}(\mathbf{k}) S^{m}(k).
\end{equation}
Here, $\tilde \phi^m(\mathbf{k})$ the structure factors $S^m(k)$ are expressed in the form of rotation-invariant tensor contractions in Eq.\ (\ref{eq:2-18-1}) and Eq.\ (\ref{eq:2-19}) respectively.
                                                                                
The density component of the solvation chemical potential in Eq.\ 
(\ref{eq:1-13}) is calculated by perturbation expansion with the
solute-solvent pair correlation function $h^{(0)}_{0s}(\mathbf{r})$
corresponding to the distribution of the solvent around the repulsive
core of the solute unaffected by the solute-solvent interaction
potential $v_{0s}$ (reference system)\cite{DMcp:93}
\begin{equation}
\label{eq:2-21}
\mu^D_{0s}=-(\beta\rho/18) \int d\mathbf{r} 
       h^{(0)}_{0s}(\mathbf{r})\left(\bm{\phi}(\mathbf{r}):\mathbf{Q}\right)^2.
\end{equation}

\section{Quadrupolar structure factors}
\label{sec:3}
The $k$-dependent quadrupolar structure factors determine the
microscopic spatial correlations of the quadrupolar polarization in
the homogeneous solvent.  Dipolar structure factors of model dipolar
fluids\cite{Fonseca:90,Raineri:93,DMjcp2:04} and fluids defined by
site-site interaction
potentials\cite{Bopp:96,Perng:96,Skaf:97,Bopp:98,Omelyan:99,Raineri:99,Perng:99}
have been rather extensively studied in the literature.  On the other
hand, calculations of the quadrupolar structure factors have never
been attempted before.  Note that both dipolar and quadrupolar
structure factors are unavailable from experiment, and liquid state
theories and computer experiment are the only source of this
information.  We present here the results of Monte Carlo (MC)
simulations of structure factors of quadrupolar hard-sphere fluids
(simulation details are given in Appendix \ref{A}).  This is followed
by an approximate analytical theory aiming at a fast algorithm
applicable to calculations of solvation thermodynamics.  In addition,
the quadrupolar structure factors for benzene used in modeling ET
reactions in Sec.\ \ref{sec:6} were obtained from Molecular Dynamics
(MD) simulations of the 12-site nonpolarizable, rigid force field
(Fig.\ \ref{fig:2}) with the quadrupole moment $Q=8.63$ D$\times$\AA{}
(simulation details are given in Appendix \ref{B}).

\begin{figure}[tbh]
 \includegraphics[width=6cm,clip=true,keepaspectratio=true]{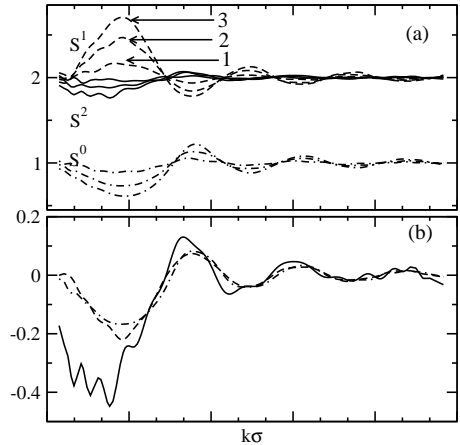}
 \caption{(a) Quadrupolar structure factors for fluids of hard spheres 
   with embedded point axial quadrupoles: $\rho^*=0.8$, $(Q^*)^2=0.1$
   (1), 0.3 (2), and 0.5 (3). (b) $(S^0(k)-1)5/8$ (dash-dotted line),
   $-(S^1(k)-2)15/32$ (dashed line), and $(S^2-2)15/8$ (solid line)
   with $S^m(k)$ from MC simulations at $(Q^*)^2=0.3$. }
\label{fig:3}
\end{figure}

The quadrupolar structure factors $S^m(k)$ of a fluid of linear
quadrupoles can be represented in terms of scalar products of unit
vectors $\mathbf{\hat e}_j$ along the principle axes of the molecule
and the direction of the wavevector $\mathbf{\hat k}=\mathbf{k}/k$.
Adopting the notation of Ref.\ \onlinecite{SPH:81},
$T_{ij}=(\mathbf{\hat e}_i\cdot \mathbf{\hat e}_j)-(\mathbf{\hat
  e}_i\cdot \mathbf{\hat k})(\mathbf{\hat e}_j\cdot \mathbf{\hat k})$
and $T_j=(\mathbf{\hat e}_j\cdot\mathbf{\hat k})$, one gets
\begin{equation}
\begin{split}
\label{eq:3-1}
S^0(k)&=\dfrac{5}{4N}\left\langle\sum_{ij} (3T_i^2-1)(3T_j^2-1)
e^{i\mathbf{k}\cdot \mathbf{r}_{ij}}\right\rangle ,\\
S^1(k)&=\dfrac{15}{N}\left\langle\sum_{ij} T_{ij} T_i T_j 
e^{i\mathbf{k}\cdot \mathbf{r}_{ij} }\right\rangle,\\
S^2(k)&=\dfrac{15}{4N}\left\langle\sum_{ij} \left[ 2T_{ij}^2-(1-T_i^2)(1-T_j^2) \right] 
e^{i\mathbf{k}\cdot \mathbf{r}_{ij}}\right\rangle .
\end{split}
\end{equation}
The axial-quadrupole structure factors $S^m(k)$ can also be expressed
in terms of projections of the solvent-solvent pair correlation
function on rotational invariants as follows
\begin{equation}
\label{eq:3-2}
\begin{split}
S^0(k)&=1-\dfrac{2}{5}\rho\left(\tilde h^{220}(k)-\tilde h^{222}(k)-4\tilde h^{224}(k) \right) ,\\
S^1(k)&=2-\dfrac{2}{5}\rho\left(2\tilde h^{220}(k)-\tilde h^{222}(k)+\dfrac{16}{3}\tilde h^{224}(k) \right) ,\\
S^2(k)&=2-\dfrac{2}{5}\rho\left(2\tilde h^{220}(k)+2\tilde h^{222}(k)-\dfrac{4}{3}\tilde h^{224}(k) \right),
\end{split}
\end{equation}                                                                                                where
\begin{equation}
\label{eq:3-2-1}
\tilde h^{22l}(k) = 4\pi i^l\int_0^{\infty}j_l(kr)h^{22l}(r)r^2dr .
\end{equation}
In Eq.\ (\ref{eq:3-2-1}), $\tilde h^{mnl}(k)$ is the Hankel transform
of the projection $h^{mnl}(r)$ of the solvent-solvent pair correlation
function on the corresponding rotational invariant,\cite{SPH:81}
$j_l(x)$ is the spherical Bessel function.\cite{Abramowitz:72}

\begin{figure}[tbh]
\includegraphics[width=6cm,clip=true,keepaspectratio=true]{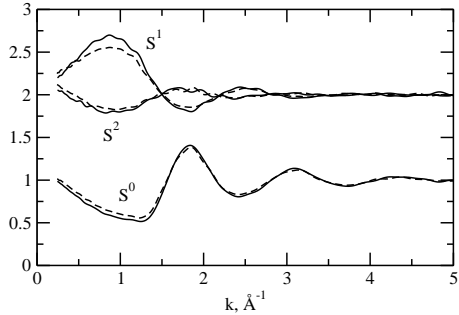}
\caption{Quadrupolar structure factors for the 12-site 
  (rigid, nonpolarizable) benzene;\cite{Danten:92} $T=298$ K (solid lines) and $T= 342$
  K (dashed lines).}
\label{fig:4}
\end{figure}

Equations (\ref{eq:3-1}) were applied to calculate $S^m(k)$ from $NVT$
MC simulations with varying quadrupole moment ($(Q^*)^2=(0.1,0.3,0.5)$
in Fig.\ \ref{fig:3}a).  It turns out that the three quadrupolar
structure factors can approximately be brought to one master curve by proper
rescaling (Fig.\ \ref{fig:3}b).  This observation indicates that
the long-range projection $\tilde h^{224}(k)$ is the main component of
$S^m(k)$ suggesting a simple perturbation approach to the calculation
of $S^m(k)$.  Taking the liquid without the quadrupolar interactions
as a reference we expand $\tilde h^{224}(k)$ in the
quadrupole-quadrupole interaction potential truncating the expansion
by the first order term.  Equation (\ref{eq:3-2}) can then be re-written
in terms of the two-particle (superscript ``(2)'') perturbation integral 
as follows
\begin{equation}
\label{eq:3-3}
\begin{split}
S^0(k) & =1-12 y_q I^{(2)}(k\sigma,\rho^*),\\
S^1(k)& = 2+16y_qI^{(2)}(k\sigma,\rho^*),\\
S^2(k) & = 2-4 y_q I^{(2)}(k\sigma,\rho^*) ,
\end{split}
\end{equation}
where
\begin{equation}
  \label{eq:4-3}
  y_q = \frac{2\pi}{5}\beta\rho Q^2/ \sigma^2
\end{equation}
is the reduced density of solvent quadrupoles\cite{DMjcp2:99} and $\sigma$
is the hard-sphere diameter. Note that the next perturbation term will
result in the three-particle perturbation integral which we do not
consider here. An improvement of the present description can be
sought in terms of a Pad\'e-truncated\cite{Gubbins:84} perturbation
expansion for $S^m(k)$.

The perturbation integral in Eq.\ (\ref{eq:3-3}) 
\begin{equation}
  \label{eq:3-4}
I^{(2)}(k\sigma,\rho^*)=\int_{1}^{\infty}dxg_{ss}^{(0)}(x,\rho^*)j_4(k\sigma x)/x^3  
\end{equation}
is defined in terms of the fourth order spherical Bessel function,
$j_{4}(x)$, and the solvent-solvent radial distribution function
$g_{ss}^{(0)}(x,\rho^*)$, where $x=r/ \sigma$. For a fluid of hard-sphere
quadrupoles, the perturbation integral depends on $k\sigma$ and the reduced
solvent density $\rho^*$. It can be approximated by a series of spherical
Bessel functions:
\begin{equation}
\label{eq:3-5}
I^{(2)}(k\sigma,\rho^*)=\sum_{n=1}^4 a_n(\rho^*) j_n(k\sigma),
\end{equation} 
where each $a_n(\rho^*)$ is a third-order polynomial in reduced density:
\begin{equation}
\label{eq:3-5-1}
a_n(\rho^*)=\sum_{p=0}^3 a_{n,p} \bigl({\rho^*}\bigr)^p.
\end{equation}
The fitting coefficients $a_{n,p}$ obtained by using the corrected
Percus-Yevick radial distribution function for hard sphere
fluids\cite{Lee:73} are listed in Table~\ref{tab:1}.  The fit
covers the range of reduced densities $\rho^*=0.5-1.0$.

\begin{table}
\caption{\label{tab:1}Third-order polynomials $a_n(\rho^*)$ in reduced density $\rho^*$ used in 
         Eqs.~(\ref{eq:3-5}) and (\ref{eq:3-5-1}). }
\begin{ruledtabular}
\begin{tabular}{ccccc}
 $p$  &  $a_{1,p}$  & $a_{2,p}$ & $a_{3,p}$ &$a_{4,p}$  \\ 
\hline
 0& 0.0095 & 0.0901  & 0.0971  & 0.0620 \\ 
 1& $-0.0776$ & 0.3580  &$-0.4647$ & 0.5574 \\ 
 2& 0.0800 & $-0.5657$  & 1.0580  &$-0.7940$ \\ 
 3& 0.0123 & 0.1290  & $-0.4370$  & 0.4810  
\end{tabular}
\end{ruledtabular}
\end{table}

\begin{figure}[tbh]
 \includegraphics*[width=6cm,clip=true,keepaspectratio=true]{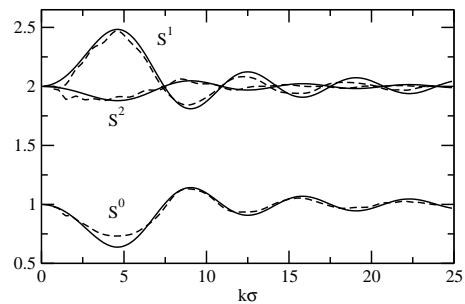}
 \caption{Quadrupolar structure factors for 
   fluids of hard-sphere axial quadrupoles from the perturbation
   expansion (Eqs.\ (\ref{eq:3-3})--(\ref{eq:3-5}), solid lines) and from the MC
   simulations (dashed lines); $\rho^*=0.8$ and $(Q^*)^2=0.3$.}
\label{fig:5}
\end{figure}

\begin{figure}

  \includegraphics*[width=6cm,clip=true,keepaspectratio=true]{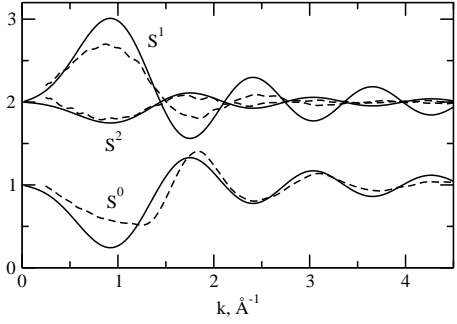}
 \caption{Quadrupolar structure factors for 
   the 12-site benzene from the perturbation expansion (Eqs.\ 
   (\ref{eq:3-3})--(\ref{eq:3-5}), solid lines) and from the MD
   simulations (dashed lines); T=298 K, $\rho^*=0.982$, and
   $(Q^*)^2=0.45$.}
\label{fig:6}
\end{figure}

The analytical equations for the structure factors [Eqs.\ 
(\ref{eq:3-3})--(\ref{eq:3-5})] are compared to MC simulation results
for the fluid of hard-sphere axial quadrupoles in
Fig.~\ref{fig:5}.  The agreement is particularly good for
$(Q^*)^2\leq0.3$. Equations (\ref{eq:3-3})--(\ref{eq:3-5}) are also used
to define the structure factors of the 12-site model of benzene (Fig.\ 
\ref{fig:2}) approximated by the axial quadrupole with the magnitude
given by Eq.\ (\ref{eq:Q}). A reasonable agreement is obtained even by
using the hard sphere radial distribution function instead of the
actual pair distribution function of the reference potential of
benzene (Fig.\ \ref{fig:6}).

\section{Model systems}
\label{sec:4}
The theory is first tested on two model solutes serving as reference
for many solvation studies: spherical ion and spherical dipole, both
dissolved in an axial-quadrupole solvent. The formal theory is tested
against the MC simulations. Two types of simulations have been carried
out.  In the first set, we obtain the axial-quadrupole structure
factors (Appendix \ref{0}) required as input to the formal theory
[Eq.\ (\ref{eq:2-20})].  In the second set, we directly calculate the
LRA chemical potential of solvation [Eq.\ (\ref{eq:1-12})] from
fluctuations of the solute-solvent potential (Appendix \ref{A}).

\subsection{Ion}
\label{sec:4-1}
For a point charge $q_0$ inside a spherical cavity one has
$\tilde\phi^{1,2}(\mathbf{k})=0$ and the only nonzero component is given by
\begin{equation}
\label{eq:4-1}
\tilde\phi^0(\mathbf{k})=\dfrac{(4 \pi q_0)^2}{5} \dfrac{j^2_1(kR_{0s})}{(k R_{0s})^2},
\end{equation}
where $R_{0s}=(\sigma_0+\sigma)/2$ is the distance of the closest
approach of the  solvent (diameter $\sigma$) to the solute (diameter $\sigma_0$).
Equations (\ref{eq:2-20})  and (\ref{eq:2-21}) then yield 
\begin{equation}
\label{eq:4-2}
-\beta\mu_{0s}=y_q(q_0^*)^2\left[I^Q(r_{0s},\rho^*,y_q) + I^D(r_{0s},\rho^*,y_q)\right],
\end{equation}
where $(q_0^*)^2 = \beta q_0^2/ \sigma$.  The perturbation integrals in Eq.\ 
(\ref{eq:4-2}) are
\begin{equation}
\label{eq:4-4}
\begin{split}
I^Q(r_{0s},\rho^*, y_q) & = \dfrac{2}{\pi r_{0s}^2}\int_0^{\infty} dx j_1^2(xr_{0s})S^0(x), \\ 
I^D(r_{0s},\rho^*, y_q) & = \int_{r_{0s}}^{\infty}\dfrac{h^{(0)}_{0s}(x)dx}{x^4}.
\end{split}
\end{equation}
The perturbation integrals depend on the solvent reduced density
$\rho^*=\rho\sigma^3$, the solvent quadrupolar density $y_q$ [Eq.\ 
(\ref{eq:4-3})], and the solute-solvent size ratio $r_{0s}=R_{0s}/ \sigma$.
They are tabulated as polynomials of $\rho^*$ and $1/r_{0s}$ in Sec.\ 
\ref{sec:4-3}.

Equation (\ref{eq:4-2}) is the microscopic perturbation solution for
solvation of an ion. Below we will also consider two approximations to
the complete solution: continuum approximation and single-particle
approximation.  The continuum limit for the solvation chemical
potential can be obtained from the microscopic formulation by assuming
that $\tilde \phi^0(k)$ changes much faster as a function of $k$ than
does the structure factor $S^0(k)$.  When this is true, one can put
$S^0(k)\simeq S^0(0)$ in Eq.~(\ref{eq:4-4}).  The density component
disappears in the continuum limit\cite{DMcp:93} with the final result
\begin{equation}
\label{eq:4-6}
-\beta \mu_{0s}^C =y_q  S^0(0) \dfrac{(q_0^*)^2}{3r_{0s}^3},
\end{equation}
where the superscript ``C'' refers to the continuum limit.  When
correlations between the solvent dipoles are neglected by assuming
$\tilde h^{22l}=0$ one gets ($S^0(0)=1$)
\begin{equation}
\label{eq:4-7}
-\beta\mu_{0s}^S =y_q\dfrac{(q^*_0)^2}{3r_{0s}^3},
\end{equation}
where the superscript ``S'' refers to the single-particle
approximation.

\subsection{Dipole}
\label{sec:4-2}
For a solute represented by a point dipole $\mathbf{m}_0$ at the
center of a spherical cavity of diameter $\sigma_0$, $\tilde\phi^{2}(\mathbf{k})=0$,
and two other components are
\begin{equation}
\begin{split}
\label{eq:4-8}
\tilde\phi^0(\mathbf{k})&=(4\pi)^2\dfrac{9(\mathbf{m}_0\cdot \mathbf{\hat k})^2}{5}\dfrac{j_2^2(kR_{0s})}{(k R_{0s}^2)^2},\\
\tilde\phi^1(\mathbf{k})&=(4\pi)^2\dfrac{3(m_0^2-(\mathbf{m}_0\cdot \mathbf{\hat k})^2)}{10}\dfrac{j_2^2(kR_{0s})}{(k R_{0s}^2)^2}.
\end{split}
\end{equation}
The total solvation chemical potential becomes
\begin{equation}
\label{eq:4-9}
-\beta\mu_{0s} = y_q(m_0^*)^2 \left[I^Q(r_{0s},\rho^*,y_q) + I^D(r_{0s},\rho^*,y_q)\right],
\end{equation}
where $(m^*_0)^2=\beta m_0^2/\sigma^3$ and the perturbation integrals are
\begin{equation}
\begin{split}
\label{eq:4-10}
I^Q(r_{0s},\rho^*,y_q) & =\dfrac{2}{\pi r_{0s}^4} \int_0^{\infty}dxj_2^2(xr_{0s})(3S^0(x)+S^1(x)), \\ 
I^D(r_{0s},\rho^*,y_q) & =5\int_{r_{0s}}^{\infty}\dfrac{h^{(0)}_{0s}(x)dx}{x^6}.
\end{split}
\end{equation}
A polynomial approximation for dipolar $I^{Q,D}(r_{0s},\rho^*,y_q)$ is given
in Sec.\ \ref{sec:4-3}.

The continuum and single-particle limits for dipolar solvation are
\begin{equation}
\label{eq:4-11}
-\beta \mu_{0s}^C  = y_q\left[3S^0(0)+S^1(0)\right] \dfrac{(m^*_0)^2}{5r_{0s}^5} 
\end{equation}
and
\begin{equation}
\label{eq:4-12}
 -\beta\mu_{0s}^S = y_q \dfrac{(m^*_0)^2}{r_{0s}^5} .
\end{equation}

        
\subsection{Perturbation integrals for $\mu_{0s}^Q$}
\label{sec:4-3}
Algebraic expressions for integrals $I^Q(r_{0s},\rho^*,y_q)$ in Eqs.\ 
(\ref{eq:4-4}) and (\ref{eq:4-10}) can be obtained by using the
perturbation expansion for the quadrupolar structure factors in Eq.\ 
(\ref{eq:3-3}). The following integrals need to be calculated:
\begin{equation}
\begin{split}
\int_0^{\infty}dx & j_1^2(x r_{0s})j_4(xy)=  \\
&\left\{ 
\begin{aligned}
\dfrac{\pi y}{1536 r_{0s}^4}(8 r_{0s}^2+y^2),\quad 2 r_{0s}\geq y \\
\dfrac{\pi }{12y^5}(5 r_{0s}^2 y^2-14 r_{0s}^4),\quad2 r_{0s}<y
\end{aligned}
\right.
\end{split}
\end{equation}
and
\begin{equation}
\begin{split}
\int_0^{\infty}dx & j_2^2(x r_{0s})j_4(xy)=\\
&\left\{ 
\begin{aligned}
\dfrac{\pi y}{15360 r_{0s}^6}(80 r_{0s}^4+30r_{0s}^2y^2-9 y^4),\quad 2 r_{0s}>y \\
\dfrac{7\pi r_{0s}^4 }{30y^5},\quad2 r_{0s}<y,
\end{aligned}
\right.
\end{split}
\end{equation}
where $x=k\sigma$ and $y=r/\sigma$.
        
The perturbation integral $I^Q(r_{0s},\rho^*,y_q)$ in Eq.\ (\ref{eq:4-4}) is then 
\begin{equation}
\label{eq:4-13}
I^Q(r_{0s},\rho^*,y_q) =  \dfrac{1}{3 r_{0s}^3}+ y_q I_{\mathrm{ion}}(r_{0s},\rho^*),
\end{equation}
where
\begin{equation}
  \label{eq:4-14}
  \begin{split}
  I_{\mathrm{ion}}(r_{0s},\rho^*) &=  28 r_{0s}^2 I_{2,8}(r_{0s},\rho^*)-10I_{2,6}(r_{0s},\rho^*)\\ 
                    & -\dfrac{1}{8r_{0s}^4}I_{1,-2}(r_{0s},\rho^*) - \dfrac{1}{64r_{0s}^6}I_{1,0}(r_{0s},\rho^*).
  \end{split}
\end{equation}
Similarly, $I^Q(r_{0s},\rho^*,y_q)$ for the dipolar solute in Eq.\ (\ref{eq:4-10}) is 
 \begin{equation}
 \label{eq:4-15}
 I^Q(r_{0s},\rho^*,y_q) = \dfrac{1}{ r_{0s}^5}+ y_q I_{\mathrm{dipole}}(r_{0s},\rho^*),
 \end{equation}
where
\begin{equation}
  \label{eq:4-16}
  \begin{split}
  I_{\mathrm{dipole}}(r_{0s},\rho^*) & = - \dfrac{28}{3} I_{2,8}(r_{0s},\rho^*)-\dfrac{5}{24r_{0s}^6}I_{1,-2}(r_{0s},\rho^*)\\
                      & -\dfrac{5}{64r_{0s}^8}I_{1,0}(r_{0s},\rho^*)-\dfrac{3}{128r_{0s}^{10}}I_{2,2}(r_{0s},\rho^*) .
  \end{split}
\end{equation}
In Eqs.\ (\ref{eq:4-14}) and (\ref{eq:4-16}), 
\begin{equation}
\label{eq:4-17}
I_{1,n}(r_{0s},\rho^*) = \int_1^{2r_{0s}}dx g_{ss}^{(0)}(x,\rho^*) x^n,
\end{equation}
and
\begin{equation}
\label{eq:4-18}
I_{2,n}(r_{0s},\rho^*) = \int_{2 r_{0s}}^{\infty}dx\dfrac{g_{ss}^{(0)}(x,\rho^*)}{x^n}.
\end{equation}
The numerical values of the perturbation integrals in Eqs.\ (\ref{eq:4-14}) and (\ref{eq:4-16})
were fit to polynomials in $\rho^*$ and $1/r_{0s}$ as follows
\begin{equation}
\label{eq:4-19}
        I_{\mathrm{ion}}(r_{0s},\rho^*)=\sum_{n=4}^{11}\dfrac{a_n(\rho^*)}{r_{0s}^{n}},
\end{equation}
\begin{equation}
\label{eq:4-20}
        I_{\mathrm{dipole}}(r_{0s},\rho^*)  =\sum_{n=6,n\neq9}^{14}\dfrac{a_n(\rho^*)}{r_{0s}^{n}},
\end{equation}
where $a_n(\rho^*)$ are third-order polynomials in $\rho^*$:
\begin{equation}
\label{eq:4-21}
     a_n(\rho^*)=\sum_{p=0}^3 a_{n,p}\bigl({\rho^*}\bigr)^p.
\end{equation}
The fit covers $\rho^*$ ranging from 0.5 to 1.0 and $r_{0s}$ ranging from
0.8 to 2.4; the coefficients $a_n$ are listed in Table \ 
\ref{tab:2}.

\begin{table*}[tbh]
\caption{\label{tab:2} Perturbation integrals from Eqs.\ 
                            (\ref{eq:4-19})--(\ref{eq:4-21}). Numbers in columns indicate the coefficients
                            $b_{np}$ in Eq.\ (\ref{eq:4-21}). }
\begin{ruledtabular}
\begin{tabular}{ccccccccc}
& \multicolumn{8}{c}{$I_{\mathrm{ion}}$} \\ 
$p$  &$a_{4,p}$&$a_{5,p}$&$a_{6,p}$&$a_{7,p}$&$a_{8,p}$&$a_{9,p}$&$a_{10,p}$&$a_{11,p}$ \\
\hline
0 & $-1/8$     &0.061    &1/64       & $-1.341$    &4.284     & $-5.611$    &3.39         & $-0.777$    \\
1 &1.618  & $-15.631$ &59.278    & $-118.0$    & 132.458 & $-83.681$ &27.214     & $-3.447$  \\
2  & $-5.240$ & 49.322   & $-190.549$ & 393.41   & $-469.0$  &323.172  & $-119.348$  & 18.254  \\
3  &3.508 & $-33.3$     &130.4       & $-274.236$ &334.8   & $-237.6$    &90.924      & $-14.506$ \\
\hline
 & \multicolumn{8}{c}{$I_{\mathrm{dipole}}$} \\
$p$ & $a_{6,p}$&$a_{7,p}$&$a_{8,p}$&$a_{10,p}$&$a_{11,p}$&$a_{12,p}$&$a_{13,p}$&$a_{14,p}$ \\
\hline
0 & $-5/24$ & $-1/8$  & 5/64 & 1/128 & 3.781 & $-8.604$ & 6.739 & $-1.8$ \\
1 & $-1.615$ & 11.195 & $-23.650$ & 87.281 & $-179.677$ & 176.884 & $-88.443$& 17.9\\
2 & 4.530 & $-32.342$ & 70.892 & $-268.563$ & 531.415 & $-492.183$ & 229.817 & $-43.5$\\
3  & $-2.614$ & 18.942 & $-42.622$ & $167.772$ & $-333.548$ & 307.841 & $-142.424$ & 26.614 \\
\hline
\end{tabular}
\end{ruledtabular}
\end{table*}

\subsection{Comparison to MC results}
\label{sec:4-4}
MC simulations of solvation of spherical ions and dipoles have been
carried out to test the formal theory. A solute is chosen as a hard sphere,
$\sigma_0/\sigma=1.8$, with charge, $(q_0^*)^2=\beta q_0^2/\sigma=15$, or with point
dipole, $(m_0^*)^2=\beta m_0^2/\sigma^3=15$, at the center. The initial
configuration in the cubic simulation box is created to accommodate
the solute at its center and $N$ solvent molecules with their size
adjusted to keep $\rho^*=0.8$. The details of reaction-field and Ewald
sum corrections for the solute-solvent and solvent-solvent interaction
potentials are given in Appendix~\ref{A}.

The results for solvation of the ionic solute are listed in
Tab.~\ref{tab:3}.  Columns 2--4 give the variance of the the
solute-solvent interaction potential from simulations. Simulations of
ion solvation at $N=256$, 500, and 864 show a noticeable size effect.
Therefore, the infinite-dilution result (column 5) was obtained by
extrapolating the data at various $N$ to $N\to\infty$.  Column 6 gives the
theoretical $\mu_{0s}$. Its separation into the quadrupolar
orientational component, $\mu_{0s}^Q$, and the density component,
$\mu_{0s}^D$, is given in columns 7 and 8, respectively. The
single-particle response $\mu_{0s}^S$ (column 9) turns out to be
surprisingly close to $\mu_{0s}$ from MC simulations.

\begin{table*}[tbh]
\caption{\label{tab:3} Formal theory and MC results for ionic solute.}
\begin{ruledtabular}
\begin{tabular}{ccccccccc}
           & $N=256$ & 500 & 864 &  $\infty$&  & &  &   \\
$(Q^*)^2$ & \multicolumn{4}{c}{$ \beta^2\langle \delta v^2_{0s}\rangle/2$\footnotemark[1]} & 
$-\beta\mu_{0s}$\footnotemark[2] &$-\beta\mu^Q_{0s}$\footnotemark[2] &$-\beta\mu^D_{0s}$ \footnotemark[2] & 
 $-\beta\mu^{S}_{0s}$\footnotemark[3]     \\
\hline
0.1            & 0.23 & 0.24 & 0.24 & 0.26 & 0.30 & 0.18 & 0.12  &  0.18 \\
0.2            & 0.42 & 0.87 & 0.44 & 0.48 & 0.60 & 0.35 & 0.25  &   0.37\\
0.3            & 0.59 & 0.60 & 0.61  & 0.65 & 0.86 & 0.49 & 0.37  &   0.55\\
0.4            & 0.73 & 0.76 & 0.77 & 0.84 & 1.12 & 0.63 & 0.49  &   0.73\\
0.5            & 0.87 & 0.89 & 0.91 & 1.00 & 1.36 & 0.74 & 0.62  &   0.92\\
0.6            & 0.99 & 1.02 & 1.04 & 1.15 & 1.60 & 0.88 & 0.72  &   1.10
\end{tabular}
\end{ruledtabular}
\footnotetext[1]{MC simulations.}
\footnotetext[2]{Eq.\ (\ref{eq:4-2}).}
\footnotetext[3]{Single-particle solution, Eq.\ (\ref{eq:4-7}). }
                                                                                                         
\end{table*}

Results for solvation of the point hard-sphere dipole are shown in
Tab.~\ref{tab:4}. No dependence of the calculated
quantities on the system size has been observed in this case for $N\geq
500$. Columns 2 and 3 give the average of the solute-solvent
interaction energy and its variance.  The near equality of these
numbers supports the use of the LRA.\cite{DMjpca:02} Column 5 gives
the orientational part $\mu_{0s}^Q$ from Eq.\ (\ref{eq:4-9}). This
component is about half of the overall solvation chemical potential as
is seen from the comparison of column 5 to columns 2 and 3. The
density component $\mu_{0s}^D$ from Eq.\ (\ref{eq:4-9}) adds to
$\mu_{0s}^Q$ to give the total $\mu_{0s}$ in column 4 which is uniformly
higher than the results of MC simulations.

In order to pin down the origin of the overestimated values, we
compare our results with the  Pad\'e form of the perturbation
expansion for a point dipole solute\cite{DMjcp2:99} (column 7). This
latter solution, which is in overall good agreement with simulations,
can also be split into the orientational and density components
(columns 8 and 9). The comparison of the orientational components of
$\mu_{0s}$ in columns 5 and 8 and the density components in columns 6
and 9 shows that it is the latter part that is overestimated in the
calculation based on Eq.\ (\ref{eq:2-21}). Finally, the
single-particle estimate, $\mu_{0s}^S$ (column 10), compares well with
$\mu_{0s}^Q$ obtained from the quadrupolar structure factors (cf.\ 
columns 5 and 10). The relatively high weight of the density component
in dipole solvation makes the single-particle approximation less
reliable than in the case of ion solvation.  However, in overall, the
single-particle formula works well for quadrupolar solvation.  This
observation is consistent with the earlier notion by Ladanyi and
Maroncelli\cite{Ladanyi:98} that the collective nature of the
solvation response diminishes for higher multipoles and that
non-dipolar solvation is dominated by single-particle response.  We
also note that the use of the quadrupolar structure factors from the
analytical equations [Eqs.\ (\ref{eq:3-3})--(\ref{eq:3-5})] gives
results nearly identical (deviation $\leq$ 1\%) to the values obtained
with simulated structure factors.

\begin{table*}[tbh]
\caption{\label{tab:4} Formal theory and MC results for dipolar solute.}
\begin{ruledtabular}
\begin{tabular}{ccccccccccc}
$(Q^*)^2$ &$-\beta\left\langle v_{0s}\right\rangle/2$\footnotemark[1] &$\beta^2\left\langle \delta v^2_{0s}\right\rangle/2$\footnotemark[1] & $-\beta\mu_{0s}$\footnotemark[2]&$-\beta\mu^Q_{0s}$\footnotemark[2] &$-\beta\mu^D_{0s}$\footnotemark[2] &$-\beta\mu_{0s}$\footnotemark[3] & $-\beta\mu^Q_{0s}$\footnotemark[3] & $-\beta\mu^D_{0s}$\footnotemark[3]  &$-\beta\mu^{S}_{0s}$\footnotemark[4]     \\
\hline
0.1            &0.59 & 0.52& 0.52&  0.26& 0.26  & 0.51& 0.27 & 0.24& 0.28 \\
0.2            & 1.07& 0.99& 1.05&  0.52& 0.53  & 0.96& 0.53& 0.44 & 0.56\\
0.3            & 1.48& 1.41& 1.56&  0.77& 0.79  & 1.38& 0.78& 0.60 & 0.84\\
0.4            & 1.91& 1.83& 2.06&  1.00& 1.06  & 1.75& 1.02& 0.73 & 1.12 \\
0.5            & 2.24& 2.19&  2.56&  1.24& 1.32  &  2.10& 1.25& 0.84 & 1.40\\
0.6            & 2.60& 2.53&  3.04&  1.46& 1.58 &  2.40& 1.46& 0.93 &  1.68
\end{tabular}
\end{ruledtabular}
\footnotetext[1]{MC simulations.}
\footnotetext[2]{Eq.\ (\ref{eq:4-9}).}
\footnotetext[3]{Pad\'e approximant from Ref.\ \onlinecite{DMjcp2:99}. The splitting into the orientational
                 and density parts of the solvation energy is achieved by putting $\rho^*=0$ in the perturbation
                 integrals when the orientational part is evaluated. }
\footnotetext[4]{Single-particle solution, Eq.\ (\ref{eq:4-12}).}
\end{table*}

\section{Comparison to experiment}
\label{sec:5}
In this section we apply the present model to the calculation of ET
rates in the donor-acceptor complex shown in Fig.\ \ref{fig:1}.
The density component of $\mu_{0s}$ is hard to estimate for a molecule
of such complex shape. On the other hand the charge in the
charge-separated state D$^+$BA$^-$ is located close to the molecular
surface. One might then expect that the solute-solvent interaction is
locally of the ion-quadrupole type. Our calculations in Sec.\ 
\ref{sec:3} show that for this type of interaction potential the
orientational part constitutes about 75\% of the overall solvation
energy. The calculations below thus assume only the orientational
component present in the solvent response
\begin{equation}
  \label{eq:5-1}
  \mu_{0s} \simeq \mu_{0s}^Q .
\end{equation}

\begin{table}[htbp]
  \centering
  \caption{\label{tab:5}Solvent parameters used in the calculations.}
\begin{ruledtabular}
  \begin{tabular}{lcccccccc}
Solvent & $\epsilon_s$ & $\epsilon_{\infty}$ & $\sigma$,\footnotemark[1] \AA{} & $\alpha$, \AA$^3$ & $m$, D & Q & 
          $\eta$ & $\epsilon_{ss}$\footnotemark[2] \\
\hline
Benzene\footnotemark[3]       & 1.00  & 1.00 & 5.28 & 10.0 & 0.  & 8.63 & 0.520 & 544 \\
Benzene       & 2.24  & 2.24 & 5.28 & 10.4 & 0.  & 8.35\footnotemark[4] & 0.515\footnotemark[1] & 544 \\
Acetonitrile  & 35.0  & 1.80 & 4.14 & 4.48 & 3.9 & 2.49\footnotemark[4] & 0.424\footnotemark[1] & 100 \\
\end{tabular}
\end{ruledtabular}
\footnotetext[1]{From Ref.\ \onlinecite{DMjpc:95}.}
\footnotetext[2]{In K, from Ref.\ \onlinecite{DMjcp1:96}.}
\footnotetext[3]{Parameters of the model benzene used in calculations.}
\footnotetext[4]{In D$\times$\AA, from Ref.\ \onlinecite{Reynolds:96}.}
\end{table}

\subsection{Free energy gap and the reorganization energy}
\label{sec:5-1}
Forward, $k_{\text{for}}(T)$, and backward, $k_{\text{back}}(T)$, rates are
available for complex \textbf{1} in benzene,\cite{Read:99,Read:00}
whereas only forward rates have been measured in
acetonitrile.\cite{Kumar:96} Rate measurements in benzene at different
temperatures give access to the overall reaction free energy
\begin{equation}
\label{eq:5-2}
 -\beta \Delta_{\mathrm{r}}G(T) = \ln\left( k_\mathrm{for}(T)\big/k_\mathrm{back}(T)\right) .
\end{equation}
The free energy gap is a composite quantity including the vacuum
energy gap $\Delta_\mathrm{vac} G$, the quadrupolar free energy $\Delta_qG$, the
free energy of solvation by induced solvent dipoles $\Delta_\mathrm{ind}
G$, and the dispersion free energy $\Delta_\mathrm{disp}G$:
\begin{equation}
\label{eq:5-3}
\Delta_\mathrm{r} G=\Delta_\mathrm{vac} G+\Delta_\mathrm{q} G+\Delta_\mathrm{ind} G+\Delta_\mathrm{disp}G.
\end{equation}
The difference in free energies is taken between the charge-separated
final state (``f'', D$^+$BA$^-$) and the initial state (``i'', D$^*$BA)
created by photoexcitation of the anthracene moiety of the
donor.\cite{Kumar:96,Read:99,Read:00} 

\begin{figure}[tbh]
  \includegraphics[width=6cm,clip=true,keepaspectratio=true]{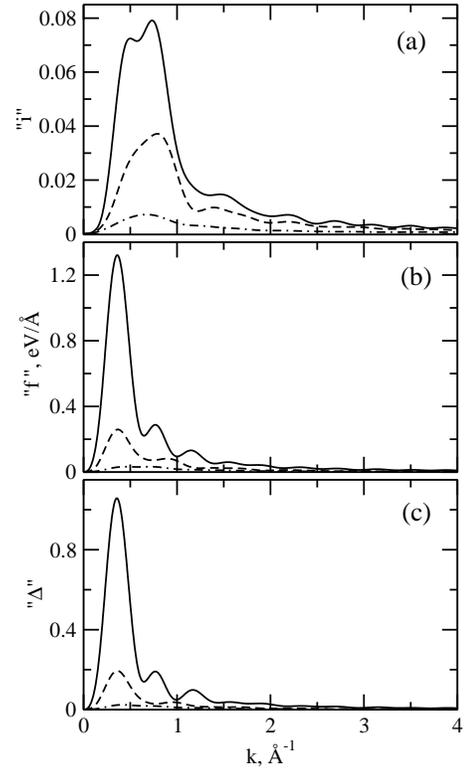}
\caption{Rotational projections $f^m_{i,f,\Delta}$ from Eqs.\ (\ref{eq:5-5-1})
  and (\ref{eq:f}) calculated for complex {\bf 1}: $m=0$ (solid
  lines), $m=1$ (dashed lines), and $m=2$ (dash-dotted lines).
  Calculations are carried out for the charge distribution in the
  initial (``i'') D$^*$BA state (a), final (``f'') D$^+$BA$^-$ state (b), and with the
  atomic charge distribution obtained as difference (``$\Delta$'') of the atomic
  charges in the final and initial states (c).}
\label{fig:7}
\end{figure}

The quadrupolar component of the energy gap is the difference of the
final and initial values of the chemical potential of solvation
\begin{equation}
\label{eq:5-4}
\Delta_\mathrm{q} G=\mu_{0s}^f-\mu_{0s}^i .
\end{equation}
Using Eq.\ (\ref{eq:2-20}),  Eq.\ (\ref{eq:5-4}) can be re-written as follows
\begin{equation}
\label{eq:5-5}
\Delta_\mathrm{q} G=-\dfrac{\beta\rho Q^2}{4\pi^2}\sum_{m}\int_0^{\infty} dk  (f_{f}^{m}(k)-f_{i}^{m}(k)) S^{m}(k),
\end{equation}
where 
\begin{equation}
  \label{eq:5-5-1}
    f^{m}_{i,f}(k)=k^2\langle \tilde\phi^{m}_{i,f}(\mathbf{k}) \rangle_{\mathbf{k}}  .
\end{equation}
The functions $f^{m}_{i,f}(k)$ are shown in Fig.\ \ref{fig:7}
(``i'' in (a) and ``f'' in (b)).

\begin{figure}[tbh]
  \includegraphics[width=6cm,clip=true,keepaspectratio=true]{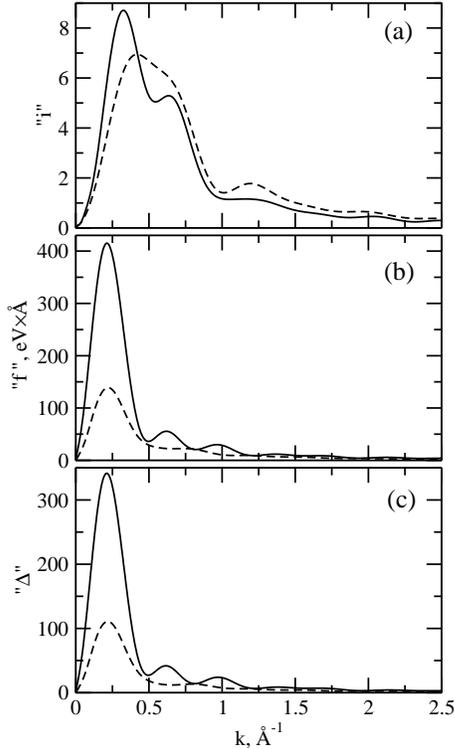}
\caption{Longitudinal (solid lines) and transverse (dashed lines) projections 
  of the electrostatic energy density [Eq.\ (\ref{eq:E})] calculated
  for complex {\bf 1}.  (a) refers to the initial (``i'') state, (b) refers to
  the final (``f'') state, (c) corresponds to the solute charge density
  obtained as difference (``$\Delta$'') in atomic charges in the final and initial
  states [Eq.\ (\ref{eq:4})].}
\label{fig:8}
\end{figure}

The induction solvation is caused by the interaction of an induced
solvent dipoles at point $\mathbf{r}$ with the electric field of the
solute. The free energy of induction solvation is obtained by
integrating the real-space electrostatic energy density
$E_{i,f}(\mathbf{r})^2$ with the distribution function
$g_{0s}(\mathbf{r})$ of the solvent molecules around the solute
\begin{equation}
  \label{eq:5-6}
  \Delta_\mathrm{ind}G = -(\rho\alpha/2) 
                  \int g_{0s}(\mathbf{r})\left[E_f(\mathbf{r})^2 - E_i(\mathbf{r})^2 \right] d\mathbf{r} .  
\end{equation}
As in the case of quadrupolar solvation, we will neglect the solute
solvent correlation function replacing $g_{0s}(\mathbf{r})$ with a
step function $\theta(\mathbf{r})$ which is equal to zero within the solute
and is equal to one otherwise. By defining the Fourier transform of
the electric field according to the relation
\begin{equation}
  \label{eq:5-7}
  \mathbf{\tilde E}(\mathbf{k}) = \int\mathbf{E}(\mathbf{r}) \theta(\mathbf{r})e^{i\mathbf{k}\cdot\mathbf{r}} d\mathbf{r}
\end{equation}
one can rewrite Eq.\ (\ref{eq:5-6}) in the form of one-dimensional $k$-integral 
\begin{equation}
\label{eq:DAG-ind}
     \Delta_\mathrm{ind}G=-\dfrac{\rho\alpha}{4\pi^2}\int_0^{\infty} dk k^2(\mathcal{E}_f(k) - \mathcal{E}_i(k)).
\end{equation}
Here, the density of electrostatic energy $\mathcal{E}(k)=\mathcal{E}^L(k)+\mathcal{E}^T(k)$ can
be separated into its longitudinal and transverse components:
\begin{equation}
\label{eq:E}
\begin{split}
\mathcal{E}^L(k) & = \langle |(\mathbf{\tilde
  E}(\mathbf{k})\cdot\mathbf{\hat k})|^2\rangle_{\mathbf{k}} \\
\mathcal{E}^T(k) & = \langle|\mathbf{\tilde
  E}(\mathbf{k})|^2\rangle_{\mathbf{k}}-\mathcal{E}^L(k). 
\end{split}
\end{equation}
The functions $k^2\mathcal{E}^{L,T}(k)$ are shown in Fig.\ 
\ref{fig:8}.

The dispersion component  $\Delta_{\text{disp}} G$ is determined as the change
in the total LJ solute-solvent interaction energy integrated over the solvent volume
$\Omega$
\begin{equation}
\label{eq:disp}
\Delta_\mathrm{disp}G=\rho\int_{\Omega} \Delta u_{LJ}(\mathbf{r})d\mathbf{r}. 
\end{equation}
Here
\begin{equation}
\label{eq:dispersion}
u_{LJ}(\mathbf{r})=
           4\sum_{a}\varepsilon_{as}\left[\left( \dfrac{\sigma_{as}}{|\mathbf{r}-\mathbf{r}_0^a|}\right)^{12} 
           -\left( \dfrac{\sigma_{as}}{|\mathbf{r}-\mathbf{r}^a_0|}\right)^{6}  \right],
\end{equation}
$\varepsilon_{as}=\sqrt{\varepsilon_0^{a}\varepsilon_{s}}$ and $\sigma_{as}=(\sigma_0^a + \sigma)/2$. The sum in Eq.\ 
(\ref{eq:dispersion}) runs over all atoms in the solute.  The atomic
diameters, $\sigma_0^a$, and LJ energies, $\varepsilon_0^a$, are parametrized with the
OPLS.\cite{OPLS} The solvent parameters are listed in Table
\ref{tab:5}.

Calculations in acetonitrile were done without the quadrupolar
component in the reaction free energy gap because of the very small
reduced quadrupole moment of acetonitrile $Q^*$ relative to its
reduced dipole $m^*$ (see Table \ref{tab:5})\cite{DMjpcb:99}
\begin{equation}
   \Delta_\mathrm{r}G=\Delta_\mathrm{vac}G+\Delta_\mathrm{p}G+\Delta_\mathrm{disp}G.
\end{equation}
The dipolar component, $\Delta_{\text{p}} G$,  is calculated from the
formalism developed in Ref.\ \onlinecite{DMjcp2:04} based on the
integration of the longitudinal, $\chi^L(k)$, and transverse,
$\chi^T(k)$, components of the dipolar response function with the
solute electric field:
\begin{equation}
  \label{eq:3}
  \Delta_{\text{p}} G = - \frac{1}{2} \int\frac{d\mathbf{k}}{(2\pi)^3} \sum_{P=L,T} \chi^P(\mathbf{k}) \mathcal{E}^P(\mathbf{k}). 
\end{equation}

The reorganization energy of polar solvation $\lambda_p$ is defined on 
the difference electric field $\mathbf{\tilde E}_{\Delta}
= \mathbf{\tilde E}_f - \mathbf{\tilde E}_i$. The electrostatic energy
density 
\begin{equation}
  \label{eq:4}
  \mathcal{E}_{\Delta}(\mathbf{k}) =\left| \mathbf{\tilde E}_{\Delta}(\mathbf{k})\right|^2 
\end{equation}
then separates into the longitudinal and transverse components
resulting in corresponding components of the reorganization energy
\begin{equation}
  \label{eq:5}
  \lambda_p = \frac{1}{2} \int \frac{d\mathbf{k}}{(2\pi)^3}  
                     \sum_{P=L,T} \chi_n^P(k) \mathcal{E}_{\Delta}^P(\mathbf{k}).
\end{equation}
In Eq.\ (\ref{eq:5}), $\chi_n^{L,T}(\mathbf{k})$ are the nuclear response
functions which, in contrast to $\chi^{L,T}(\mathbf{k})$, do not contain
the effect of induced solvent dipoles.\cite{DMjcp2:04,DMcp:05} 

The reorganization energy of quadrupolar solvation $\lambda_q$ is given in
terms of the gradient of $\mathbf{E}_{\Delta}(\mathbf{k})$. Following our
formal derivation in terms of spherical coordinates, the final result
can be written as
\begin{equation}
\label{eq:lambda-benzene}
\lambda_q = \dfrac{\beta\rho Q^2}{4\pi^2}\sum_{m}\int dk f_{\Delta}^{m}(k) S^{m}(k),
\end{equation} 
where 
\begin{equation}
\label{eq:f}
   f_{\Delta}^{m}(k) =  k^2\langle \tilde\phi^{m}_{\Delta}(\mathbf{k}) \rangle_{\mathbf{k}} 
\end{equation}
and $\tilde \phi_{\Delta}^{m}(\mathbf{k})$ is obtained from the solute charge
distribution build on difference atomic charges in the final and
initial states (Fig.\ \ref{fig:7}c).

\subsection{ET rates}
\label{sec:5-2}
The calculations according to the formalism outlined above require the
combination of two input components, from the solute and from the
solvent. The solute requires specifying coordinates, van der Waals
radii (OPLS parameterization\cite{OPLS}), and charges of the solute
atoms. The volume accessible to the solvent is limited by the solvent
accessible surface (SAS) defined by complementing the solute atomic
radii with the half of the solvent hard-sphere diameter, $(\sigma_0^a + \sigma)/2$.
Atomic coordinates and charges for complex \textbf{1} (Fig.\ 
\ref{fig:1}) are from Ref.\ \onlinecite{Troisi:04}. The Fourier
transforms of the solute field and solute field gradient are carried
out on the 256$^3$ grid as described in Appendix \ref{C}.
 
The calculation of the dispersion component $\Delta_{\text{disp}}G$
requires knowledge of the alteration of the solute LJ energy with
electronic transition.  Since this information is not available, we
use the London dispersion potential to connect the change in the LJ
energy to the change in the dipolar polarizability.  The initial ET
state is experimentally produced by photoexcitation of the anthracene
moiety of the donor.  Anthracene polarizabilities in the ground
(subscript ``g'') and excited (subscript ``e'') states are\cite{Renge:92} $\alpha_g=25$
\AA$^3$ and $\alpha_e = 42$ \AA$^3$. According to the London
equation $\varepsilon^a_{g/e} \sim \alpha_{g/e}^2$ (``a'' refers to an atomic site
within the solute).  The LJ energies on the anthracene atoms might
hence be expected to scale as $\varepsilon_{e}^a=(\alpha_e/\alpha_g)^2\varepsilon_{g}^a$ leading to
the scaling of the solute-solvent dispersion interaction potential as
$\alpha_e/\alpha_g$.  In the calculations below we will assume that LJ energies
on only the anthracene moiety change with the transition.  Since the
anion acceptor state of complex \textbf{1} might involve some unknown
polarizability change off-setting the polarizability change of the
anthracene moiety, we consider $\Delta\alpha = \alpha_e-\alpha_g$ as a fitting parameter
to reproduce the experimental $\Delta_r G(T)$ [Eq.\ (\ref{eq:5-2})]. The
other fitting parameter is $\Delta_{\text{vac}} G$.

\begin{widetext}
\begin{table*}[tbh]
\caption{\label{tab:6} Thermodynamics parameters (eV) of 
equilibrium solvation of complex \textbf{1} in benzene.}
\begin{ruledtabular}
\begin{tabular}{ccccccccccc}
 $T$/K & $\Delta_\mathrm{q} G\footnotemark[1]$ &  $\Delta_\mathrm{q} G$\footnotemark[2]  
       & $\Delta_\mathrm{ind} G$ & $\Delta_\mathrm{disp} G$\footnotemark[3] &  $\Delta_\mathrm{disp} G$\footnotemark[4] &
         $\Delta_\mathrm{r} G$\footnotemark[5]&
         $\Delta_\mathrm{r} G$\footnotemark[6] & $\Delta_\mathrm{r} G\footnotemark[7]$ & $\lambda_q$\footnotemark[1] & $\lambda_q$\footnotemark[2] \\
\hline
  298 & $-0.244$ & $-0.236$  &$-0.322$ & 0.196& 0.460& $-0.113$ &$-0.106$ & $-0.110$\footnotemark[8] & 0.205 & 0.198\\
  312 & $-0.226$ & $-0.221$  &$-0.317$ & 0.193& 0.453& $-0.092$ &$-0.090$ & $-0.097$                 & 0.189 & 0.185\\
  326 & $-0.218$ & $-0.208$  &$-0.312$ & 0.190& 0.446& $-0.082$ &$-0.085$ & $-0.082$                 & 0.183 & 0.174\\
  342 & $-0.203$ & $-0.196$  &$-0.307$ & 0.186& 0.438& $-0.066$ &$-0.072$ & $-0.065$                 & 0.170 & 0.164\\
\end{tabular}
\end{ruledtabular}
\footnotetext[1]{Eqs.\ (\ref{eq:5-5}) and (\ref{eq:lambda-benzene}) with $S^{0,1,2}(k)$ from MD simulations.}
\footnotetext[2]{Eqs.\ (\ref{eq:5-5}) and (\ref{eq:lambda-benzene}) with $S^{0,1,2}(k)$ 
                 from Eqs.\ (\ref{eq:3-3}) and (\ref{eq:3-5}).}
\footnotetext[3]{Calculated with $\Delta \alpha=7.22$ \AA$^3$.}
\footnotetext[4]{Calculated with $\Delta \alpha=17$ \AA$^3$.}
\footnotetext[5]{Calculated with $\Delta \alpha=7.22$ \AA$^3$ and $\Delta_\mathrm{vac}G=0.258$ eV.}
\footnotetext[6]{Calculated with $\Delta \alpha=17$ \AA$^3$ and $\Delta_\mathrm{vac}G=-0.68\times10^{-3}$ eV.}
\footnotetext[7]{Experiment from Eq.\ (\ref{eq:5-2}).}
\footnotetext[8]{Corresponds to T=296 K.}
\end{table*}
\end{widetext}              

The second input to our calculation formalism comes from properties of
the pure solvent. The polar, quadrupolar and dipolar, response comes
into the theory in the form of three quadrupolar structure factors
$S^m(k)$ and two dipolar structure factors $S^{L,T}(k)$. A
parameterization scheme based on solvent dipole moment $m$, solvent
diameter $\sigma$, solvent polarizability $\alpha$, and solvent number density
$\rho$ was developed previously in Refs.\ \onlinecite{DMjcp2:04} and
\onlinecite{DMcp:05}. Solvent parameters used in the calculation are
listed in Table \ref{tab:5}.

The results of calculation of $\Delta_r G(T)$ and $\lambda_q(T)$ in benzene are
listed in Tab.~\ref{tab:6}.  Data in columns 2 and 10 in
Table \ref{tab:6} have been calculated by using $S^m(k)$
from MD simulations (12-site benzene).  Equations (\ref{eq:3-3}) and
(\ref{eq:3-5}) provide an alternative analytical route to $S^m(k)$
through the effective linear quadrupole moment with its magnitude
given by Eq.\ (\ref{eq:Q}).  Results of using the analytical structure
factors in Eqs.~(\ref{eq:5-5}) and (\ref{eq:lambda-benzene}) agree
well with corresponding values obtained by using the structure factors
from MD simulations (cf.\ columns 2, 3 and 10, 11 in
Table~\ref{tab:6}).

The reaction entropy $\Delta S^0=-d\Delta_\mathrm{r}G/dT$ is equal to
$-8.55\,\mathrm{k_B}$  when $\Delta \alpha=17$ \AA{} of anthracene is used
to calculate $\Delta_\mathrm{disp}G$ (column 6). The fit of experimental $\Delta_rG(T)$
then yields $\Delta_\mathrm{vac}G=-0.68 \times 10^{-3}$ eV.  For these
parameters, the dispersion and induction solvation entropies almost
cancel each other, $\Delta_\mathrm{ind} S^0=-4.06\,\mathrm{k_B}$ (column 4)
and $\Delta_\mathrm{disp} S^0=5.80\,\mathrm{k_B}$ (column 6), making the
quadrupolar solvation entropy, $\Delta_\mathrm{q} S^0=-10.29\,\mathrm{k_B}$
(column 2), nearly equal to the experimental reaction entropy.  The
experimental value $\Delta S^0= - 11.84\,\mathrm{k_B}$ (column 7) can be
reproduced by downward scaling of $\Delta \alpha$ to $7.22$ \AA$^3$ leading to
$\Delta_\mathrm{vac}G=0.258$ eV. The dispersion solvation entropy then
scales down to $\Delta_\mathrm{disp} S^0=2.46\,\mathrm{k_B}$ (column 5).

\begin{table}[tbh]
\caption{\label{tab:7} Rates (10$^8$ s$^{-1}$) of the ET in benzene.}
\begin{ruledtabular}
\begin{tabular}{ccccccc}
T/K&$k_\mathrm{for}$\footnotemark[1]&$k_\mathrm{for}$\footnotemark[2] &$k_\mathrm{for}$\footnotemark[3] &$k_\mathrm{back}$\footnotemark[1] &$k_\mathrm{back}$\footnotemark[2] &$k_\mathrm{back}$\footnotemark[3] \\
\hline
298 & 29.8\footnotemark[4] & 28.54&25.07& 0.48\footnotemark[4]  & 0.32&0.36 \\
312 & 27.5 & 26.86& 24.64 & 0.75& 0.79&0.79 \\
326 & 24.6 & 26.02& 24.97 & 1.31& 1.25 &1.11 \\
342 & 21.5 & 24.52& 24.78 & 2.41& 2.45 &1.97
\end{tabular}
\end{ruledtabular}
\footnotetext[1]{Experiment from Ref.\ \onlinecite{Read:00}.}
\footnotetext[2]{Theory with $\Delta \alpha$=7.22 \AA$^3$, $\Delta_{\text{vac}}G=0.258$ eV, and $|V|=8.0$ cm$^{-1}$.}
\footnotetext[3]{Theory with $\Delta \alpha$=17 \AA$^3$, $\Delta_{\text{vac}}G=-0.68\times 10^{-3}$ eV, and $|V|=7.75$ cm$^{-1}$.}
\footnotetext[4]{Measurement at  $T=296$ K.}
\end{table}
                                                                                                         
\begin{figure}[tbh]  
  \includegraphics[width=6cm,clip=true,keepaspectratio=true]{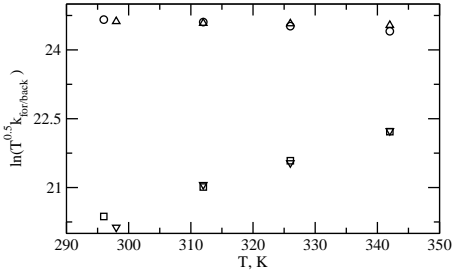}
   \caption{Forward (circles and up triangles) and backward (squares and down triangles) ET rates 
     in benzene for complex {\bf 1}. Triangle symbols correspond to
     experimental data.\cite{Read:00} Circles and squares correspond
     to four temperatures at which $S^{0,1,2}(k)$ have been calculated
     from MD simulations. }
   \label{fig:9}
\end{figure}

The parameters $\Delta \alpha$ and $\Delta_{\text{vac}}G$ from the fit of
experimental $\Delta_rG(T)$ have been used to calculate ET rates in benzene
(Tab.~\ref{tab:7}) according to the Jornter-Bixon formula
(nonadiabatic ET):
\begin{eqnarray}
\label{eq:5-20}
  k_\mathrm{for/back}(T)=\dfrac{|V|^2}{\hbar} \left(\dfrac{\pi \beta}{\lambda_s}\right)^{1/2}\sum_{n=0}^{\infty}e^{-S}\dfrac{S^n}{n!}\\\nonumber
 \exp\left( -\dfrac{\beta(\lambda_s(T)\pm\Delta_r G(T)+nh\nu_v)^2}{4\lambda_s}\right) .
\end{eqnarray}
Here, $S=\lambda_v/h\nu_v$ is the Huang-Rhys factor and ``+'' and ``$-$'' refer
to the forward (``for'') and backward (``back'') reaction rates,
respectively. The vibrational reorganization energy $\lambda_v=0.39$ eV and
vibrational frequency $h\nu_v=1410$ cm$^{-1}$ are from Ref.\ 
\onlinecite{Read:00}. The experimental rates $k_{\text{for/back}}(T)$
are reproduced with the electronic coupling $|V|=8$ cm$^{-1}$ and $\Delta_r
G(T)$ and $\lambda_q(T)$ from our calculations (columns 3 and 6 in
Tab.~\ref{tab:7}, see also Fig.\ \ref{fig:9}).

The parameters $\Delta \alpha$ and $\Delta_\mathrm{vac}G$ obtained from the fit of
$\Delta_r G(T)$ in benzene refer to the solute in the gas phase and do not
depend on the solvent. The reliability of our fitting procedure can
thus be further tested by using these parameters to calculate ET rates
in a different solvent. Since we also want a test independent of the
present formulation for quadrupolar solvation, acetonitrile with its
small quadrupole moment and large dipole moment presents an ideal
choice.  Table \ref{tab:8} lists the calculated thermodynamic
parameters and rates of charge separation in complex {\bf 1} in
acetonitrile in the experimental temperature range between 255 and 335
K.\cite{Kumar:96} The calculations of the free energy gap and the
reorganization energy are done by using the recently developed
microscopic theory of dipolar solvation.\cite{DMjcp2:04,DMcp:05} The
dispersion part of the ET driving force in acetonitrile is calculated
according to Eqs.(\ref{eq:disp}) and (\ref{eq:dispersion}) with
the solvent parameters listed in Table \ref{tab:5}.

\begin{table}[tbh]
\caption{\label{tab:8} Free energies (eV) and rate constants ($10^8$ s$^{-1}$) for 
                             complex \textbf{1} in acetonitrile.}
\begin{ruledtabular}
\begin{tabular}{cccccc}
T/K & $\Delta_\mathrm{p}G$ & $\Delta_\mathrm{disp}G$ & $\Delta_\mathrm{r}G$  & $\lambda_p$ 
    & $k_\mathrm{for}$ \\
\hline
255 & $-1.739$ & 0.101 & $-1.381$ & 1.48 &  2.91\\
265 & $-1.688$ & 0.099 & $-1.331$ & 1.41 &  3.28\\
275 & $-1.640$ & 0.098 & $-1.285$ & 1.35 &  3.66\\
285 & $-1.597$ & 0.096 & $-1.243$ & 1.30 &  4.03\\
295 & $-1.557$ & 0.095 & $-1.204$ & 1.25 &  4.39\\
305 & $-1.513$ & 0.094 & $-1.161$ & 1.20 &  4.64\\
315 & $-1.485$ & 0.092 & $-1.134$ & 1.16 &  5.07\\
325 & $-1.452$ & 0.091 & $-1.103$ & 1.12 &  5.40\\
335 & $-1.422$ & 0.090 & $-1.074$ & 1.08 &  5.70
\end{tabular}
\end{ruledtabular}
\end{table}

\begin{figure}[tbh]
   \includegraphics[width=6cm,clip=true,keepaspectratio=true]{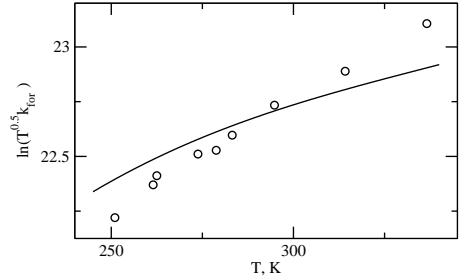}
   \caption{The forward rates of charge separation in complex {\bf 1} in acetonitrile. 
     Open circles refer to experiment, the solid line refers to
     theory.}
   \label{fig:10}
\end{figure}

Since the electronic coupling may depend on the
solvent\cite{ZimmtWaldeck:03} the experimental reaction
rates\cite{Kumar:96} were fit to Eq.\ (\ref{eq:5-20}) with electronic
coupling $V$ considered as the only fitting parameter (Fig.\
\ref{fig:10}).  This procedure results in $|V|=2.21$ cm$^{-1}$.
This value is almost 4 times smaller than the corresponding electronic
coupling in benzene.  This trend parallels the one reported by Zimmt
and Waldeck:\cite{ZimmtWaldeck:03} $V=7.2$ cm$^{-1}$ in benzene vs
$V=4.6$ cm$^{-1}$ in acetonitrile.  Qualitatively, this difference is
attributed to the higher overlap of the donor and acceptor orbitals of
complex {\bf 1} with the molecular orbitals of benzene residing in the
clamp compared to the molecular orbitals of
acetonitrile.\cite{ZimmtWaldeck:03}

\section{Discussion}
\label{sec:6}
Theories of dipolar solvation have been developed over the last
decades to provide a hierarchy of approximations ranging from the
simple estimates of the Born and Onsager equations to more accurate
Poisson-Boltzmann equation\cite{delphi} and up to microscopic
liquid-state calculations in terms of perturbation
expansions,\cite{DMjcp1:99,DMjcp2:99} integral
equations,\cite{Raineri:99} or density
functional\cite{Ramirez:02,Ramirez:05} theories.  Nothing of
comparable scale exists for quadrupolar solvation.  Kim and
co-workers\cite{Jeon:01,Dorairaj:02,Kim1:03,Kim2:03} have formulated a
continuum theory of quadrupolar solvation of spherical solutes. The
theory provides a fast calculation formalism, but suffers from the
approximation of the spherical solute shape and empirical
parameterization of the quadrupolar solvent susceptibility [Eq.\ 
(\ref{eq:1-7})]. The integral equation theories of the RISM
family\cite{Raineri:99,Yamaguchi:02} avoid the problem of separate
calculations of dipolar and quadrupolar solvation by considering the
site-site interaction potentials automatically incorporating the
infinite-order multipole expansion.  This advantage comes at the
expense of the necessity to solve rather complex integral equations on
one hand and to supply the charge-charge structure factors of the
solvent characterized by its site-site force field on the other.
Force field parameterization is available only for a few solvents, and
the calculations of the charge-charge structure factors requires
extensive computations. Therefore, one would want a formalism for
quadrupolar solvation that would incorporate the quadrupole moment of
the solvent along with some experimentally available parameters as
input in the microscopic solvation formalism. The formulation of such
a theory is the result of this paper. It turns out that the
quadrupolar structure factors of non-dipolar solvents can indeed be
rather accurately calculated based on experimental input parameters
for the solvent: quadrupole moment, density, and the effective
molecular diameter.  Based on these structure factors we have
formulated a hierarchy of approximations of increasing sophistication
and accuracy listed in Table \ref{tab:9}.

\begin{widetext}
\begin{table*}
\caption{Hierarchy of solutions for the solvation 
         chemical potential, $-\beta\mu_{0s}^Q$, in quadrupolar solvents.}
\label{tab:9}
\begin{ruledtabular}
  \begin{tabular}{lll}
Approximation &  \multicolumn{2}{c}{Solute} \\
       & Ion & Dipole \\
\hline
Single-particle & $(q_0^*)^2y_q/3r_{0s}^3$       &  $(m_0^*)^2y_q/r_{0s}^5$ \\
Continuum       & $(q_0^*)^2y_qS^0(0)/3r_{0s}^3$ &  $(m_0^*)^2y_q(3S^0(0)+S^1(0))/5r_{0s}^5$ \\ 
Perturbation    & $(q_0^*)^2y_q \bigl(1/3r_{0s}^3+y_qI_\mathrm{ion}(r_{0s},\rho^*)\bigr)$ &     $(m_0^*)^2y_q  \bigl(1/r_{0s}^5+y_qI_\mathrm{dipole}(r_{0s},\rho^*)\bigr) $ \\
\hline
     &     \multicolumn{2}{c}{Arbitrary solute} \\
\hline
Perturbation & \multicolumn{2}{l}{$(\beta^2\rho Q^2/2) \sum_m \int \tilde\phi^m(\mathbf{k}) S^m(k) d\mathbf{k}/(2\pi)^3$ }\\
\end{tabular}
\end{ruledtabular}
\end{table*}
\end{widetext}

As the crudest estimate we offer the single-particle approximation
which allows one to calculate the quadrupolar solvation energy from
two solvent parameters, the quadrupolar density $y_q$ [Eq.\ 
(\ref{eq:4-3})] and the solvent diameter $\sigma$. The continuum
approximation includes $S^m(0)$.  However, since $S^0(0)\simeq 1$ and
$S^{1,2}(0)\simeq2$, the continuum limit is in fact very close to the
single-particle result. The next approximation is based on
perturbation integrals depending on the reduced solvent density
$\rho^*=\rho\sigma^3$ and the reduced distance of the closest solute-solvent
approach $r_{0s}=\sigma_0/2\sigma + 0.5$.  The number of solvent parameters thus
rises to three: $y_q$, $\sigma$, and $\rho^*$.  Finally, the calculation of
solvation free energy of a solute of arbitrary shape requires three
angular projections of the auto correlation function of quadrupolar
polarization (last line in Table \ref{tab:9}).  The problem which is
still remain unresolved is the absence of an accurate algorithm for
the density component of the solvation chemical potential $\mu_{0s}^D$
[Eqs.\ (\ref{eq:1-12})]. This component can be included for spherical
solutes since the pair solute-solvent spherically symmetric
distribution function can be then computed with sufficient accuracy.
For instance, in the case of a spherical dipole, the Pad\'e form of
perturbation theories\cite{DMjcp2:99} gives a good agreement with
simulations.

\begin{figure}[htbp]
  \centering
  \includegraphics*[width=6cm]{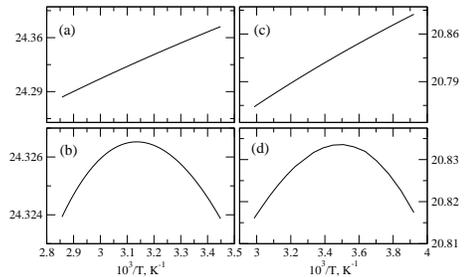}

  \caption{$\ln(k_\mathrm{for})$ vs $1/T$ for complex \textbf{1} is calculated in benzene with $\Delta_{\text{vac}}G=0.173\,\mathrm{eV}$ and in acetonitrile with $\Delta_{\text{vac}}G=0.19\,\mathrm{eV}$. In benzene $\lambda_q$ and $\Delta_{\mathrm{r}}G$: (a) are fixed at $T=320$ K; (b) assume full temperature dependence. In acetonitrile $\lambda_q$ and $\Delta_{\mathrm{r}}G$: (c) are fixed at $T=275$ K; (d) assume full temperature dependence. }
  \label{fig:11}
\end{figure}

Several effective medium approximations have been proposed in the
literature to deal with the problem of the local density
profile.\cite{YamaguchiJPC:02} They replace the solute-solvent
interaction potential in the orientational component of the chemical
potential of solvation ($v_{0s}\theta(\mathbf{r})$ in Eq.\ (\ref{eq:1-12}))
with an effective solute-solvent coupling including the information
about the solute-solvent density correlations (direct solute-solvent
correlation function in the
surrogate\cite{Raineri:99,Yamaguchi:02,YamaguchiJPC:02} and density
functional\cite{BagchiBiswas:99} theories or vertex in the
mode-coupling formulation\cite{Egorov:02}). However, all such
approximations are heavily dependent on the detailed knowledge of the
solute-solvent distribution function which is available only for
solutes of simple shape.  There is still no acceptable theory
adequately addressing the calculation of the $\mu_{0s}^D$ components for
solutes of complex shape.

The present calculation for complex \textbf{1} highlights the
importance of the induction and dispersion components of solvation in
the equilibrium energy gap of ET reactions (Table
\ref{tab:6}). The contribution of induction and
dispersion forces to the solvent reorganization energy are given,
however, by a higher order of the perturbation theory, and they are
normally much smaller than the corresponding components of the free
energy gap.\cite{DMmp:95,DMjcp:95} Quadrupolar solvation then becomes
the most significant contribution to the solvent reorganization energy
in non-dipolar solvents. In the crudest approximation, the quadrupolar
reorganization energy is proportional to the quadrupolar density
$y_q$: 
\begin{equation}
  \label{eq:6-2}
  \lambda_q \propto q_0^2\rho\beta Q^2/(R_0 + \sigma/2)^3
\end{equation}
for an ion and
\begin{equation}
  \label{eq:6-3}
  \lambda_q \propto m_0^2\rho\beta Q^2/(R_0 + \sigma/2)^5
\end{equation}
for a dipole.  The present theory thus predicts a negative slope of
$\lambda_q$ vs $T$:
\begin{equation}
 \label{eq:6-4}
    \left(\partial \lambda_q/\partial T\right)_P \simeq - \lambda_q(T^{-1} +\alpha_P) , 
\end{equation}
where $\alpha_P$ is the isobaric volume expansion coefficient. The slope of
$\lambda_q$ vs $P$ is positive and is proportional to isobaric
compressibility $\beta_T$:
\begin{equation}
\label{eq:6-5}
  \left(\partial \lambda_q/\partial P\right)_T \simeq \beta_T \lambda_q . 
\end{equation}

The negative slope of the solvent reorganization energy has been
obtained here for both quadrupolar (Table \ref{tab:6})
and dipolar (Table \ref{tab:8}) solvents. This property thus
seems to be a universal signature of ET reorganization in polar
(dipolar and quadrupolar) solvents. This particular dependence of the
reorganization energy on temperature results in an observable effect
in reactions with low activation barrier as was first noted in Ref.\ 
\onlinecite{DMmp:93}. The classical Marcus activation barrier for ET
passes through a maximum when plotted vs $1/T$ (Arrhenius plot) at the
point of activationless ET $\lambda_{q,p}(T)+\Delta_{\text{r}} G(T) = 0$ when the
negative slope of $\lambda_{q,p}$ vs $T$ is applied. Figure \ref{fig:11}
illustrates this point by showing $\ln(k_\mathrm{for})$ vs $1/T$
calculated on complex \text{1} for which $\Delta_{\text{vac}}G$ was
adjusted to allow the point $\lambda_{q,p}(T)+\Delta_{\text{r}} G(T) = 0$ to fall
in the experimental range of temperatures. The calculations with full
temperature dependent $\lambda_{q,p}(T)$ and $\Delta_{\text{r}} G(T)$ are
compared to the those under assumption $\lambda_{q,p}=Const$, $\Delta_{\text{r}}
G=Const$. No maximum is seen in the latter case. We note also that
intramolecular vibrations tend to mask the appearance of the maximum.

\begin{figure}[htbp]
  \centering
  \includegraphics*[width=8cm]{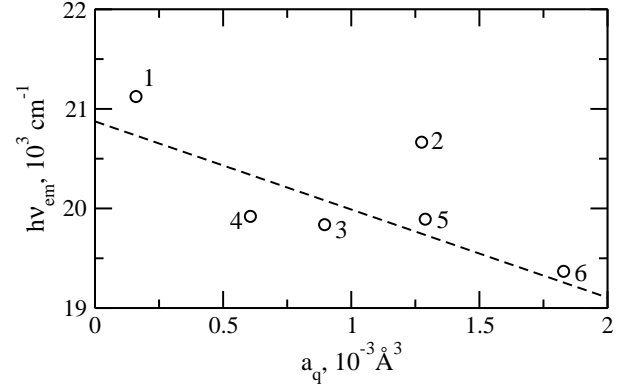}
  \caption{Emission energy of ADMA\cite{Khajehpour:00} vs $a_q$ [Eq.\ (\ref{eq:6-7})] 
    in quadrupolar solvents: ethylene (1), CO$_2$ (2), benzene (3),
    tetrafluorobenzene (4), toluene (5), 1,4-dioxane (5). Emission
    energies, solvent parameters, and $R_0=4.32$ \AA{} are taken from
    Ref.\ \onlinecite{Khajehpour:00}. The dashed line is drawn as regression through the solvents
    excluding CO$_2$. }
  \label{fig:12}
\end{figure}

Although the full version of the theory is preferable for the analysis of
spectroscopy and ET kinetics, the simple single-particle approximation is
useful in analyzing the qualitative trends with changing non-dipolar solvent.
In particular, quadrupolar solvation results in the solvatochromic shift of
optical transitions resulting in the change of chromophore's dipole state
$\mathbf{m}_i \to \mathbf{m}_f$ 
\begin{equation}
  \label{eq:6-6}
  h\Delta\nu_{i\to f} = - 2 a_q(\mathbf{m}_f -\mathbf{m}_i)\cdot\mathbf{m}_i ,
\end{equation}
where in the single-particle approximation,
\begin{equation}
  \label{eq:6-7}
  a_q = \frac{2\pi\beta\rho Q^2}{5(R_0 + \sigma /2)^5} .
\end{equation}
Figure \ref{fig:12} illustrates this trend based on the emission
energies of ADMA reported by Khajehpour and
Kauffman.\cite{Khajehpour:00} Note that emission frequencies are
affected by dispersion and induction solvation and the deviation of
the supercritical CO$_2$ from the linear trend may be traced to its
lower polarizability resulting in lower dispersion and induction
stabilization energies.

The present formulation, combined with the previously developed model
for dipolar solvation,\cite{DMjcp2:04,DMcp:05} covers two extreme
cases of polar solvation -- purely dipolar and purely quadrupolar
solvents. The formalism is based on the correlation functions of
dipolar and quadrupolar polarization as input and, therefore, can be
applied to an arbitrary dipolar or quadrupolar solvent. This approach
has allowed us to address the problem of microscopic calculation of
solvation thermodynamics for solutes with atomic resolution of
geometry and charge distribution.  The application to real large
donor-acceptor molecules shows encouraging results. Generalization of
the theory will require considering dipolar-quadrupolar solvents,
which will be a subject of future work.

\begin{acknowledgments}
  Acknowledgments are made to the donors of The
  Petroleum Research Fund, administered by the ACS (39539-AC6)
  for support of this research. We are grateful to Dr.\ A.\ Troisi
  for sharing with us structural data on the cleft molecule and to
  Prof.\ M.\ B. Zimmt and Prof.\ D.\ H.\ Waldeck for the experimental 
  kinetic data.
\end{acknowledgments}  

\appendix
\section{{\label{0}}Derivation of Eq.\ \ref{eq:2-20}.}
Here we provide the derivation of the solute field gradient and
quadrupolar structure factors in terms of rotationally-invariant
contraction of Cartesian tensors.  The Cartesian components of
$Q_{2m}$ and $\phi_{2m}$ are
\begin{equation}
\label{eq:2-15}  
\begin{split}
Q_{20}=&\left( \dfrac{5}{4\pi}\right)^{1/2}Q_{zz},\\
Q_{21}=&-\sqrt{ \dfrac{5}{6\pi}}\left( Q_{xz}+iQ_{yz}\right) ,\\
Q_{22}=&\sqrt{ \dfrac{5}{24\pi} }\left( Q_{xx}-Q_{yy}+2iQ_{xy}\right),
\end{split}
\end{equation}
and
\begin{equation}
  \label{eq:2-16}
  \begin{split}
   \phi_{20} & = -\sqrt{\frac{\pi}{5}}\phi_{zz},\\
   \phi_{21} & = \sqrt{\frac{4\pi}{30}}\left(\phi_{xz}+i\phi_{yz}\right),\\
   \phi_{22} & =  -\sqrt{\frac{\pi}{30}}\left(\phi_{xx} - \phi_{yy}+2i\phi_{xy}\right) .
  \end{split}
\end{equation}
Also, for negative $\underline{m}=-m$, $Q_{2,\underline{m}}=(-1)^m
Q^*_{2m}$, $\phi_{2,\underline{m}}=(-1)^m \phi^*_{2m}$.  For the structure
factors in the $X'Y'Z'$ coordinates ($\mathbf{k}$ is collinear to
$Z'$, Fig.\ \ref{fig:2}) one has
%
\begin{equation}
\begin{split}
\label{eq:2-17}
S'^{0}(k)&=\dfrac{5}{NQ^2}\left\langle \sum_{i,j} Q'_{zz,i}Q'_{zz,j}e^{i \mathbf{k}\cdot \mathbf{r}_{ij}} \right\rangle,\\
S'^{1}(k)&=\dfrac{20}{3NQ^2}\left\langle \sum_{i,j} \left( Q'_{xz,i}Q'_{xz,j}+Q'_{yz,i}Q'_{yz,j}\right)  e^{i \mathbf{k} \cdot\mathbf{r}_{ij} } \right\rangle,\\
S'^{2}(k)&=\dfrac{5}{3NQ^2}\biggl\langle \sum_{i,j} \bigl( (Q'_{xx,i}-Q'_{yy,i})(Q'_{xx,j}-Q'_{yy,j})+\\
      & 4Q'_{xy,i}Q'_{xy,j} \bigr) 
      e^{ i \mathbf{k} \cdot\mathbf{r}_{ij}} \biggr \rangle.
\end{split}
\end{equation}
Since $\mathbf{\hat k}'=(0,0,1)$ in $X'Y'Z'$ coordinates
and $\mathbf{Q}'_{\alpha\beta}$ is a symmetric and traceless tensors one gets
%
\begin{equation}
\begin{split}
\label{eq:2-18}
  Q'_{zz,i} & Q'_{zz,j}  =\left( \mathbf{\hat k}' \cdot\mathbf{Q}_i' \cdot \mathbf{\hat k}'\right) 
                 \left( \mathbf{\hat k}' \cdot \mathbf{Q}_j' \cdot \mathbf{\hat k}'\right),\\
  Q'_{xz,i} & Q'_{xz,j}+Q'_{yz,i}Q'_{yz,j} =\left( \mathbf{\hat k}' \cdot \mathbf{Q}_i' \cdot \mathbf{Q}_j'\cdot \mathbf{\hat k}'\right) -\\
        & \left( \mathbf{\hat k}' \cdot \mathbf{Q}_i' \cdot \mathbf{\hat k}'\right) 
    \left( \mathbf{\hat k}'\cdot \mathbf{Q}_j'\cdot \mathbf{\hat k}'\right),\\
  (Q'_{xx,i} &  -Q'_{yy,i})(Q'_{xx,j}-Q'_{yy,j})+4Q'_{xy,i}Q'_{xy,j}  =2\mathbf{Q}'_{i}:\mathbf{Q}'_{j}- \\
        & 4\left( \mathbf{\hat k}' \cdot\mathbf{Q}_i'\cdot
    \mathbf{Q}_j'\cdot\mathbf{\hat k}'\right)+\left( \mathbf{\hat k}'\cdot
    \mathbf{Q}_i' \cdot \mathbf{\hat k}'\right) \left( \mathbf{\hat
      k}'\cdot \mathbf{Q}_j'\cdot \mathbf{\hat k}'\right) .
\end{split}
\end{equation}
The tensor contractions in Eq.\ (\ref{eq:2-18}) are invariant under
rotations of the coordinate system. Therefore Eq.\ (\ref{eq:2-17}) can be rewritten as

\begin{equation}
\begin{split}
\label{eq:2-19}
S^0(k)&=\dfrac{5}{NQ^2}\left\langle\sum_{i,j}
\left( \mathbf{\hat k}\cdot \mathbf{Q}_i\cdot \mathbf{\hat k}\right)  
\left( \mathbf{\hat k}\cdot \mathbf{Q}_j \cdot \mathbf{\hat k}\right) e^{i \mathbf{k}\cdot \mathbf{r}_{ij}}\right\rangle, \\
S^1(k)&=\dfrac{20}{3NQ^2}
\biggl\langle\sum_{i,j}
\biggl[
\left( \mathbf{\hat k}\cdot \mathbf{Q}_i\cdot\mathbf{Q}_j \cdot\mathbf{\hat k}\right)\\
    &- \left( \mathbf{\hat k}\cdot \mathbf{Q}_i \cdot \mathbf{\hat k}\right)
\left( \mathbf{\hat k}\cdot \mathbf{Q}_j \cdot \mathbf{\hat k}\right)
\biggr]e^{i \mathbf{k}\cdot \mathbf{r}_{ij}}\biggr\rangle,\\
S^2(k)&=\dfrac{5}{3NQ^2}\biggl\langle\sum_{i,j}
\biggl[2\mathbf{Q}_i:\mathbf{Q}_j
-4\left( \mathbf{\hat k} \cdot \mathbf{Q}_i \cdot\mathbf{Q}_j\cdot \mathbf{\hat k}\right) \\
 & + \left( \mathbf{\hat k}\cdot \mathbf{Q}_i \cdot \mathbf{\hat k}\right)
\left( \mathbf{\hat k}\cdot \mathbf{Q}_j \cdot \mathbf{\hat k}\right)
\biggr]e^{ i \mathbf{k}\cdot \mathbf{r}_{ij} }
\biggr\rangle,
\end{split}
\end{equation}
where $\mathbf{\hat k}= \mathbf{k}/\left| \mathbf{k} \right|$. Also the spherical components $\tilde \phi^m(\mathbf{k})$ entering Eq.\ (\ref{eq:2-20}) 
can be given in the rotation-invariant form as
\begin{equation}
\begin{split}
\label{eq:2-18-1}
\tilde \phi^0(\mathbf{k})&=\frac{1}{20}|(\mathbf{\hat k} \cdot \bm{\phi}\cdot   \mathbf{\hat k})|^2,\\
\tilde \phi^1(\mathbf{k})&=\dfrac{1}{30}\left((\mathbf{\hat k}\cdot \bm{\phi}^*\cdot \bm{\phi}
  \cdot
  \mathbf{\hat k})-|(\mathbf{\hat k} \cdot \bm{\phi }\cdot   \mathbf{\hat k})|^2\right),\\
\tilde \phi^2(\mathbf{k})&=\dfrac{1}{120}\left(2\bm{\phi}^*:\bm{\phi}-4(\mathbf{\hat
    k}\cdot \bm{ \phi}^*\cdot \bm{\phi} \cdot \mathbf{\hat k})+|(\mathbf{\hat k} \cdot
  \bm{\phi}\cdot \mathbf{\hat k})|^2\right) ,
\end{split}
\end{equation}
where $\bm{\phi}$ stands for $\tilde \phi_{\alpha\beta}(\mathbf{k})$. A numerical
algorithm for the calculation of $\tilde \phi_{\alpha\beta}(\mathbf{k})$ used in
the calculations of complex \textbf{1} is outlined in Appendix\ 
\ref{C}.

\section{\label{A}Details of MC simulations}
\subsection{Quadrupolar hard-sphere fluids}
MC simulations have been carried out on a system of $N=500$ hard
sphere molecules with axial quadrupoles placed in a cubic simulation
box. Periodic boundary conditions, minimum image convention, and the
ratio of 0.5 between the distance of interaction cutoff and the size
of the simulation box have been adopted.\cite{Allen:96} The cutoff of
the quadrupole-quadrupole interactions is corrected by the reaction
field of continuum dielectric with the dielectric constant $\varepsilon'=\infty$.
Simulation runs of the average length of $9 \times10^5$ cycles with $4
\times10^5$ cycles used to calculate $S^m(k)$ were performed at $(Q^*)^2=\beta\rho
Q^2/\sigma^5=(0.1,0.2,0.3,0.4,0.5,0.6)$ and $\rho^*=\rho\sigma^3=0.8$.

\subsection{Dipole solute in quadrupolar liquid}
We use the reaction field (RF) correction for the cutoff of long-range
electrostatic interactions in MC simulations. The interaction energy
of a dipolar solute with the quadrupolar solvent is the sum of the
interaction energy $u_{0s}^{DQ}(j)$ with the quadrupoles residing within
the cutoff sphere and the RF correction terms
$u_{0s}^{DQ;\mathrm{RF}}$ and $u_{0s}^{DD;\mathrm{self}}$
\begin{equation}
\label{eq:b2-1}
u_{0}=\sum_j \left[u_{0s}^{DQ}(j) + u_{0s}^{DQ;\mathrm{RF}}(j)\right] + u_{0s}^{DD;\mathrm{self}}.
\end{equation}
where
\begin{equation}
  \label{eq:b2-2}
  u_{0s}^{DQ}(j) =r_{j0}^{-4} \left[ \left( \mathbf{\hat r}_{j0}\cdot \mathbf{Q}_j\cdot\mathbf{\hat r}_{j0} \right) \left( \mathbf{m}_0  \cdot\mathbf{\hat r}_{j0} \right) - 2 \left( \mathbf{\hat r}_{j0}\cdot \mathbf{Q}_j\cdot  \mathbf{m}_{0} \right) \right] ,
\end{equation}
and $\mathbf{r}_{j0}=\mathbf{r}_{j}-\mathbf{r}_{0}$, $\mathbf{\hat
  r}_{j0}=\mathbf{r}_{j0}/ r_{j0}$.  In Eq.\ (\ref{eq:b2-1}),
$u_{0s}^{DQ;\mathrm{RF}}$ is the interaction energy of the solute with
the polarization of the dielectric continuum (dielectric constant
$\epsilon'$) outside the cutoff sphere with the radius $R_c$
\begin{equation}
  \label{eq:b2-3}
u_{0s}^{DQ;\mathrm{RF}}(j)=\frac{6(\mathbf{r}_{j0}\cdot \mathbf{Q}_j\cdot  \mathbf{m}_{0})}{R_c^5}\, \dfrac{\varepsilon'-1}{3\varepsilon'+2}  .
\end{equation}
The term $u_{0s}^{DD;\mathrm{self}}$ in Eq.\ (\ref{eq:b2-1}) is the
energy of solute self-polarization by the dielectric outside the
cutoff:
\begin{equation}
\label{eq:b2-4}
u_{0s}^{DD;\mathrm{self}} = -\dfrac{m_0^2}{R_c^3}\dfrac{\varepsilon'-1}{2\varepsilon'+1} .
\end{equation}

The energy of the $j$th solvent molecule in the RF geometry is
\begin{equation}
\label{eq:b2-5}
u_{ss}(j)=\sum_{m\neq j} \left[ u_{ss}^{QQ}(jm)+ u_{ss}^{QQ;\mathrm{RF}}(jm)\right] + u_{ss}^{QQ;\mathrm{self}}(j) + u_{s0}^{QD}(j) .
\end{equation}
The quadrupole-quadrupole interaction energy is
\begin{equation}
  \label{eq:b2-6}
\begin{split}
u_{ss}^{QQ}(jm)&=\dfrac{1}{3r_{mj}^5}\left(35\bigl( 
\mathbf{\hat r}_{mj}\cdot \mathbf{Q}_m\cdot  \mathbf{\hat r}_{mj} \right)\times \\
    & \times \left( \mathbf{\hat r}_{mj}\cdot \mathbf{Q}_m\cdot \mathbf{\hat r}_{mj} \right)- 
       20 \left( \mathbf{\hat r}_{mj}\cdot \mathbf{Q}_j\cdot \mathbf{Q}_m\cdot \mathbf{\hat r}_{mj} \right)\\ 
    & +2 \left( \mathbf{Q}_{j}: \mathbf{Q}_m\right)\bigr) , 
\end{split}  
\end{equation}
where $\mathbf{r}_{mj}=\mathbf{r}_m-\mathbf{r}_j$, $\mathbf{\hat r}_{mj}=\mathbf{r}_{mj}/r_{mj}$, and
$\mathbf{Q}_j:\mathbf{Q}_m=\sum_{\nu,\alpha} Q_{j,\nu\alpha}Q_{m,\nu\alpha}$.  The interaction
energy of the $j$th solvent quadrupole with the polarization of the
dielectric continuum outside the cavity induced by the $m$th ($m\neq j$)
solvent molecule is
\begin{equation}
  \label{eq:b2:7}
u_{ss}^{QQ;\mathrm{RF}}(jm)=-\dfrac{ 2 \left( \mathbf{Q}_{j}: \mathbf{Q}_m  \right) }{R_c^5}\dfrac{\varepsilon'-1}{3\varepsilon'+2} .
\end{equation}
The self energy of the $j$th  quadrupole is
\begin{equation}
  \label{eq:b2-8}
u_{ss}^{QQ;\mathrm{self}}(j) =-\dfrac{\left( \mathbf{Q}_{j}: \mathbf{Q}_j\right) }{R_c^5}\dfrac{\varepsilon'-1}{3\varepsilon'+2} .
\end{equation}
Finally, $u_{s0}^{QD}(j)$ in Eq.\ (\ref{eq:b2-5}) is equal to $u_{0s}^{DQ}(j)$
from Eq.\ (\ref{eq:b2-1}).

For an axial quadrupole, $\mathbf{Q}_{\nu\alpha,j}=(Q/2)\left( 3\mathbf{\hat
    e}_{\nu,j}\mathbf{\hat e }_{\alpha,j} -\delta_{\nu\alpha} \right)$, where $\delta_{\nu\alpha}$ is
the Kronecker delta symbol. The interaction potentials are given
by
\begin{equation}
\label{eq:aB-5}
\begin{split}
u_{0s}^{DQ} (j)& = \dfrac{3m_0Q}{2r_{j0}^4}\biggl( 
5\left(\mathbf{\hat e}_j\cdot \mathbf{\hat r}_{j0}\right)^2 \left(\mathbf{\hat e}_0 \cdot\mathbf{\hat r}_{j0}\right)\\
       &-2\left(\mathbf{\hat e}_j\cdot \mathbf{\hat r}_{j0}\right)\left(\mathbf{\hat e}_0\cdot \mathbf{\hat e}_j\right)-\left(\mathbf{\hat e}_0 \cdot\mathbf{\hat r}_{j0}\right) \biggr) ,\\
u_{ss}^{QQ}(jm) & =  \dfrac{3Q^2}{4r_{mj}^5}\biggl(1 - 
5\left( \mathbf{\hat e}_j\cdot \mathbf{\hat r}_{mj}\right)^2-5\left( \mathbf{\hat e}_m\cdot  \mathbf{\hat r}_{mj}\right)^2\\
        & -20\left( \mathbf{\hat e}_j\cdot  \mathbf{\hat e}_m\right)\left( \mathbf{\hat e}_m\cdot  \mathbf{\hat r}_{mj}\right)
        \left( \mathbf{\hat e}_j\cdot  \mathbf{\hat r}_{mj}\right) \\
        & +35\left( \mathbf{\hat e}_j\cdot  \mathbf{\hat r}_{mj}\right)^2
        \left( \mathbf{\hat e}_m\cdot  \mathbf{\hat r}_{mj}\right)^2\biggr) ,\\
u_{ss}^{QQ;\mathrm{RF}}(jm)&=- \dfrac{3Q^2}{2R_c^5}\dfrac{\varepsilon'-1}{3\varepsilon'+2}
               \left( 3\left(\mathbf{\hat e}_j\cdot  \mathbf{\hat e}_m\right)^2-1\right) ,\\
u_{ss}^{QQ;\mathrm{self}}(j) & =-\dfrac{3Q^2}{2R_c^5}\dfrac{\varepsilon'-1}{3\varepsilon'+2} .\\
\end{split}
\end{equation}

\subsection{Ionic solute in quadrupolar liquid}
For ion solvation simulations we use the RF approximation for the
solvent-solvent interactions [Eq.\ (\ref{eq:aB-5})] and the Ewald sum
(ES) approximation for the solute-solvent interactions.  The solute
energy is 
\begin{equation}
        u_{0s}=\sum_{j=1}^N u_{0s}^{CQ}(j) + u_{0s}^{CC; \mathrm{self}},
\end{equation}
where $u_{0s}^{CC; \mathrm{self}}$ is the self energy term and $
u_{s0}^{QC}(j)$ is the interaction energy of the solute with the $j$th
solvent molecule. The self energy is the interaction energy of the
solute charge with its images in replicas of the simulation box
\begin{equation}
u_{0s}^{CC; \mathrm{self}}=-\dfrac{q_0^2}{L}\left( \dfrac{\xi_{EW}}{2}+
          \dfrac{\pi}{6}\left( \dfrac{\sigma_0}{L}\right)^2  - 
          \dfrac{\pi^2}{180}\left( \dfrac{\sigma_0}{L}\right)^5\right),
\end{equation}
where the finite-size correction $\xi_{EW}=1.41865$ is according to
Hunenberger and McCammon,\cite{McCammon:99} $\sigma_0$ is the hard-sphere
solute diameter, and $L$ is the side length of the cubic simulation
box.

The interaction energy of the solute with the $j$th solvent molecules
is ($q_0$ is the solute charge)
\begin{equation}
\label{eq:B1}
\begin{split}
u_{0s}^{CQ}(j)&= \dfrac{q_0}{3}\bigl( \mathbf{\hat r}_{j0}\cdot\mathbf{Q}_j\cdot\mathbf{\hat r}_{j0} \bigr)
 \left[ f_2(r_{j0})+\dfrac{3}{r_{j0}^3}f_1(r_{j0})\right]\\
 &-\dfrac{4q_0\pi}{3L^3}\sum_{|\mathbf{k}|\neq0}\exp(-k^2/4\kappa^2)\cos(\mathbf{k}\cdot\mathbf{r}_{j0})
      \bigl( \mathbf{\hat k}\cdot\mathbf{Q}_j\cdot\mathbf{\hat k} \bigr),
\end{split}
\end{equation}
 where
\begin{equation}
\begin{split}
f_1(x)&=\mathrm{erfc}(\kappa x)+\dfrac{2\kappa x\exp(-x^2\kappa^2)}{\sqrt{\pi}} , \\
f_2(x)&=\dfrac{4\kappa^3\exp(-x^2\kappa^2)}{\sqrt{\pi}} .
\end{split}
\end{equation}
and $\mathrm{erfc}(x) = 1 -\mathrm{erf}(x)$, $\mathrm{erf}(x)$ is the
error function.\cite{Abramowitz:72} The solute-solvent interaction
potential consists of two parts. The first is taken in the real
$\mathbf{r}$-space, while the second is taken in the inverted
$\mathbf{k}$-space. The convergence parameter $\kappa$ defines the number
of replicas of the simulation cell to be taken in the real space.
Additionally, it defines the number of $\mathbf{k}$-vectors taken in
the inverted space sum. When this parameter is sufficiently large the
calculation the real space sum is restricted to the original
simulation box.\cite{Allen:96}

Finally, the interaction energy of the $j$th molecule of the solvent is 
\begin{equation}
u_{s}(j)=u_{ss}^{QQ}(j)+u_{ss}^{QQ;RF}(j)+u_{ss}^{QQ;\mathrm{self}}(j)+u_{s0}^{QC}(j),
\end{equation}
where the first three terms are given by Eq.\ (\ref{eq:aB-5}),
$ u_{s0}^{QC}(j) = u_{0s}^{CQ}(j)$ [Eq.\ (\ref{eq:B1})], and the
last term is given by Eq.\ (\ref{eq:B1}).

\section{{\label{B}}Details of MD simulations of benzene}
The MD simulations were carried out with the force field of 12-site
benzene by Danten \textit{et al}.\cite{Danten:92} (Table\ \ref{tab:10}).
The site-site interaction is given by the sum of the Lennard-Jones
(LJ) and Coulomb interaction potentials:
\begin{equation}
E^{ab}=4\varepsilon^{ab}\left[\left(\dfrac{\sigma^{ab}}{r^{ab}}\right)^{12}-\left(\dfrac{\sigma^{ab}}{r^{ab}}\right)^{6}\right]
  +\dfrac{q^{a}q^{b}}{r^{ab}} ,
\end{equation}
where the LJ parameters are taken according to the Lorentz-Bertholet
rules: $\varepsilon^{ab}=\sqrt{\varepsilon^{a}\varepsilon^{b}}$ and $\sigma^{ab}=(\sigma^{a}+\sigma^{b})/2$. All
simulations were done with the DL\_POLY molecular dynamics
package.\cite{dlpoly20:96} We run MD simulations in the temperature
range from 298 K to 342 K with a 14 K step. The timestep in each
simulation is 5 fs. All MD simulation are 10 ns long.

We used the Nos\'e-Hoover\cite{hoover:85} thermostat with the relaxation
parameter 0.5 fs. The proper choice of the simulation timestep and the
relaxation parameter ensures low drift in the total energy of about
0.3\%. Cut-off distance of 12 \AA\, is used in calculations of LJ
interactions. Ewald summation parameters used in calculations are: (1)
the convergence parameter $\kappa=0.265$ \AA$^{-1}$; (2) the maximum
wavelength $k_{x,y,z}^\mathrm{max}=7$ \AA$^{-1}$.  The simulation box is
constructed to include 125 benzene molecules in a cube with the side
length $L=26.46$ \AA\, at $T=298$ K to reproduce the experimental mass
density of benzene,\cite{crc:85ed} $\rho_M=0.874$ g/cm$^3$. The side
length is adjusted at each temperature to give the correct
experimental value for the isobaric temperature expansion
coefficient,\cite{crc:85ed} $\alpha_p=1.14\times10^{-3}$\,K$^{-1}$.

\begin{table}[tbh]
 \caption{\label{tab:10} Force field of benzene used in the MD simulation.}
 \begin{ruledtabular}
 \begin{tabular}{cccc}
  Interaction site\footnotemark[1] & $\sigma^{a}/$\AA& $\epsilon^{a}\times10^{2}$/(kcal/mol)& $q^{a}/e$ \\  \hline
  C& 3.473 & 83.1& $-0.153$ \\
  H & 2.945 & 12.5&  0.153
 \end{tabular}
 \end{ruledtabular}
 \footnotetext[1]{$r_{\mathrm{CC}}=1.393$ \AA, $r_{\mathrm{CH}}=1.027$ \AA.}
 \end{table}

\section{{\label{C}}Calculation algorithm for $\tilde\phi_{\alpha\beta}(\mathbf{k})$}

\begin{figure}[tbh] 
  \includegraphics[width=7cm]{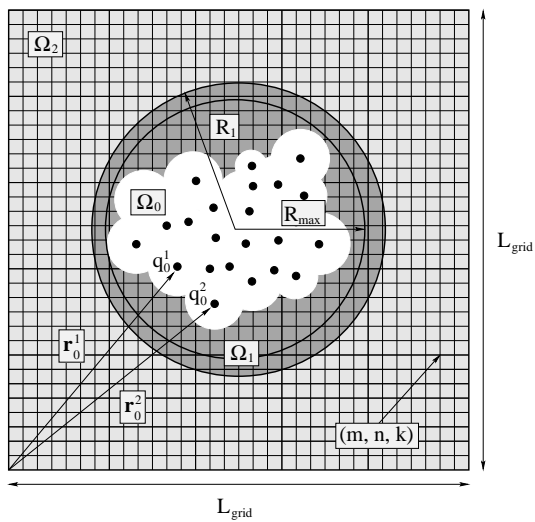}
   \caption{The schematic representation of the calculation geometry. 
     $\Omega_0$ is the solvent inaccessible cavity.  $\Omega_1$ (space between
     the sphere of radius $R_1$ and $\Omega_0$) is the region of the
     numerical Fourier transform of $\phi_{\alpha\beta}(\mathbf{r})$ [Eq.\ 
     (\ref{eq:2-2})]. $\Omega_2$ is the region where Fourier transform of
     $\phi_{\alpha\beta}(\mathbf{r})$ is taken analytically [Eq.\ (\ref{eq:C2})].
     $R_\mathrm{max}$ is the radius of the sphere defining solute
     maximum extension and $R_1=R_{\text{max}} + \sigma/2$. The length
     $L_\mathrm{grid}=9\times R_\mathrm{max}$ is the size of the
     $\mathbf{r}$-space grid of dimension
     $\mathrm{256}\times\mathrm{256}\times\mathrm{256}$; $(m,n,k)$ is the
     position of a point on the grid.}
   \label{fig:13}
\end{figure}

The gradient of the solute electric field $\phi_{\alpha\beta}(\mathbf{r})$ is
calculated on the 256$^3$ grid in $\mathbf{r}$-space (Fig.\ 
\ref{fig:13}) as defined by Eq.\ (\ref{eq:2-2}). In order to speed up
the calculations, we split the calculation of the Fourier transform
$\tilde\phi_{\alpha\beta}(\mathbf{k})$ into two regions. In region $\Omega_1$, the
Fourier transform is calculated numerically with a fast Fourier
transform (FFT) algorithm.\cite{Fortran:96} In region $\Omega_2$, the
multipole expansion of the set of solute charges is used, and the
Fourier transform is calculated analytically.  The total
$\tilde\phi_{\alpha\beta}(\mathbf{k})$ is the sum of Fourier
transforms from each region
\begin{equation}
\tilde \phi_{\alpha\beta}(\mathbf{k})=\tilde \phi_{\alpha\beta;\,\Omega_1}(\mathbf{k})+\tilde\phi_{\alpha\beta;\,\Omega_2}(\mathbf{k}) .
\label{eq:C1}
\end{equation}

The analytical part of Fourier transform for an arbitrary distribution
of solute atomic charges $q_0^a$ with coordinates $\mathbf{r}_0^a$, $r_0^a < R_1$ is
given by
\begin{eqnarray}
\label{eq:C2}
\begin{split}
\tilde\phi_{\alpha\beta;\,\Omega_2} (\mathbf{k}) & = -4\pi \sum_{a=1}^M\ q_0^a \sum_{n=2}^{\infty}(-i)^n\left(\dfrac{r_0^a}{R_1}\right)^{n-2}
\dfrac{j_{n-1}(kR_1)}{kR_1}\\ 
     & \times \bigl( \hat r^a_{0\alpha}\hat r^a_{0\beta}P_{n-2}''(x^a)-\delta_{\alpha\beta}P'_{n-1}(x^a)\\
     & - (\hat k_{\alpha}\hat r^a_{0\beta}+\hat r^a_{0\alpha}\hat k_{\beta})P_{n-1}''(x^a) +\hat k_{\alpha}\hat k_{\beta} P_{n}''(x^a)\bigr) .
\end{split}
\end{eqnarray}
Here, $R_1$ is the radius of the sphere enclosing volume $\Omega_1$,
$j_n(x)$ is the spherical Bessel function of the order
$n$,\cite{Abramowitz:72} $P_n(x)$ is the Legendre polynomial of 
order $n$, $P_n'(x)=dP_n(x)/dx$ and $P_n''(x)=d^2P_n(x)/dx^2$,
$x^a=\mathbf{\hat k}\cdot\mathbf{\hat r}_0^a$.

The volume of integration for the numerical FFT is a cubic box of side
lenght
\begin{equation}
\label{eq:C3}
L_\mathrm{grid}=f \times R_\mathrm{max}
\end{equation}
obtained as a multiple of the maximum extension of the solvent
inaccessible cavity, given by the radius $R_\mathrm{max}$.  The radius
of the sphere of the solute maximum extension is defined as
\begin{equation}
\label{eq:C4}
   R_\mathrm{max}
  =\mathrm{max}\left\lbrace\left|\mathbf{r}_0^a - \mathbf{r}_\mathrm{gc}\right|+\sigma_0^a/2\right\rbrace ,
\end{equation}
where $i$ runs over $M$ solute atoms. The sphere is centered at the
geometric center of the solute molecule,
$\mathbf{r}_\mathrm{gc}=(1/M)\sum_a\mathbf{r}_0^a$ and $\sigma_0^a$ is the
diameter of atom $a$ of the solute. The radius of the spherical region
$\Omega_1$ is the obtained by adding the solvent radius $\sigma/2$ to
$R_{\text{max}}$: $R_1=R_{\text{max}} + \sigma/2$ (Fig.\ \ref{fig:13}). The
multiplication factor $f=9$ in Eq.\ (\ref{eq:C3}) is chosen as a
trade-off between minimizing FFT errors from artificial periodicity of
the lattice sum and the need for a sufficiently small increment in the
$\mathbf{k}$-space for the $\mathbf{k}$-integration.

Charges of complex $\mathbf{1}$ in the non-polar (initial) and
charge-transfer (final) states are calculated with
$\mathtt{Gaussian'03}$ (UHF/6-31G(3p)).\cite{G:03} Partial atomic
charges are obtained by fitting to the electrostatic potential given
by the exact wavefunction in the CHELPG scheme.\cite{Breneman:90}
Charges are calculated separately for the isolated donor, donor
cation, acceptor, acceptor anion, and the neutral bridge. The charges
of the two hydrogens that substitute the rest of the clamp in the
model compounds are incorporated in the connected carbon
atoms.\cite{Troisi:04}

\bibliographystyle{apsrev}
\bibliography{/home/dmitry/p/bib/chem_abbr,/home/dmitry/p/bib/et,/home/dmitry/p/bib/etnonlin,/home/dmitry/p/bib/liquids,/home/dmitry/p/bib/solvation,/home/dmitry/p/bib/dm,/home/dmitry/p/bib/dynamics,/home/dmitry/p/bib/glass,/home/dmitry/p/bib/lc,/tmp/qsolv/miscQ}

\begin{thebibliography}{88}
\bibitem[{com()}]{comQ:1}
\bibinfo{note}{The term non-polar solvation is often associated with all
  situations when either the solute or the solvent do not posses both charge
  and dipole moment. We limit ourselves here with a more narrow case of an
  ionic or dipolar solute in a non-dipolar solvent.}
\end{thebibliography}

\end{document}